%
%
\magnification=1200\overfullrule=0pt\baselineskip=15pt
\vsize=22truecm \hsize=15truecm \overfullrule=0pt
\pageno=0
\font\smallfont=cmr8 scaled \magstep0
\font\ismallfont=cmti9 scaled \magstep0
\font\titlefont=cmbx10 scaled \magstep2
\font\ititlefon=cmbxti10 scaled \magstep2
\font\sectnfont=cmbx10 scaled \magstep1
\font\isectnfon=cmbxti10 scaled \magstep1
\font\mamfont=msbm10 scaled \magstep0
\font\tenmsa=msam10
\font\sevenmsa=msam7
\font\fivemsa=msam5
\newfam\msafam
\textfont\msafam=\tenmsa  \scriptfont\msafam=\sevenmsa
  \scriptscriptfont\msafam=\fivemsa
\def\hexnumber@#1{\ifcase#1 0\or1\or2\or3\or4\or5\or6\or7\or8\or9\or
        A\or B\or C\or D\or E\or F\fi }
\edef\msa@{\hexnumber@\msafam}
\mathchardef\lesssim="3\msa@2E
\mathchardef\gtrsim="3\msa@26
\font\vsf=cmr6 scaled \magstep0
\def\build#1_#2{\mathrel{\mathop{\kern 0pt#1}\limits_{#2}}}
\def\mname{\ifcase\month\or January \or February \or March \or April
           \or May \or June \or July \or August \or September
           \or October \or November \or December \fi}
\def\date{\hbox{\strut\mname \number\year}}
\def\scnum{\hbox{SPhT-96/001\strut}}
\def\banner{\hfill\hbox{\vbox{\offinterlineskip
                               \scnum\date}}\relax}
\footline={\ifnum\pageno=0{}\else\hfil\number\pageno\hfil\fi}
%
%
%
\newcount\FIGURENUMBER\FIGURENUMBER=0
\def\fig#1{\expandafter\ifx\csname FG#1\endcsname\relax
               \global\advance\FIGURENUMBER by 1
               \expandafter\xdef\csname FG#1\endcsname
                      {\the\FIGURENUMBER}\fi
           Fig.~\csname FG#1\endcsname\relax}
\def\figand#1#2{\expandafter\ifx\csname FG#1\endcsname\relax
               \global\advance\FIGURENUMBER by 1
               \expandafter\xdef\csname FG#1\endcsname
                      {\the\FIGURENUMBER}\fi
           \expandafter\ifx\csname FG#2\endcsname\relax
               \global\advance\FIGURENUMBER by 1
               \expandafter\xdef\csname FG#2\endcsname
                      {\the\FIGURENUMBER}\fi
           Figs.~\csname FG#1\endcsname\ and
                   \csname FG#2\endcsname\relax}
\newcount\TABLENUMBER\TABLENUMBER=0
\def\table#1{\expandafter\ifx\csname TB#1\endcsname\relax
               \global\advance\TABLENUMBER by 1
               \expandafter\xdef\csname TB#1\endcsname{\the\TABLENUMBER}\fi
             Table \csname TB#1\endcsname\relax}
\newcount\REFERENCENUMBER\REFERENCENUMBER=0
\def\reftag#1{\expandafter\ifx\csname RF#1\endcsname\relax
               \global\advance\REFERENCENUMBER by 1
               \expandafter\xdef\csname RF#1\endcsname
                      {\the\REFERENCENUMBER}\fi
             }
\def\ref#1{\expandafter\ifx\csname RF#1\endcsname\relax
               \global\advance\REFERENCENUMBER by 1
               \expandafter\xdef\csname RF#1\endcsname
                      {\the\REFERENCENUMBER}\fi
             [\csname RF#1\endcsname]\relax}
\def\refto#1#2{\expandafter\ifx\csname RF#1\endcsname\relax
               \global\advance\REFERENCENUMBER by 1
               \expandafter\xdef\csname RF#1\endcsname
                      {\the\REFERENCENUMBER}\fi
           \expandafter\ifx\csname RF#2\endcsname\relax
               \global\advance\REFERENCENUMBER by 1
               \expandafter\xdef\csname RF#2\endcsname
                      {\the\REFERENCENUMBER}\fi
             [\csname RF#1\endcsname--\csname RF#2\endcsname]\relax}
\def\refand#1#2{\expandafter\ifx\csname RF#1\endcsname\relax
               \global\advance\REFERENCENUMBER by 1
               \expandafter\xdef\csname RF#1\endcsname
                      {\the\REFERENCENUMBER}\fi
           \expandafter\ifx\csname RF#2\endcsname\relax
               \global\advance\REFERENCENUMBER by 1
               \expandafter\xdef\csname RF#2\endcsname
                      {\the\REFERENCENUMBER}\fi
            [\csname RF#1\endcsname,\csname RF#2\endcsname]\relax}
\newcount\EQUATIONNUMBER\EQUATIONNUMBER=0
\def\EQNO#1{\expandafter\ifx\csname EQ#1\endcsname\relax
               \global\advance\EQUATIONNUMBER by 1
               \expandafter\xdef\csname EQ#1\endcsname
                      {\the\EQUATIONNUMBER}\fi
            \eqno(\csname EQ#1\endcsname)
                      \relax}
\def\eq#1{\expandafter\ifx\csname EQ#1\endcsname\relax
               \global\advance\EQUATIONNUMBER by 1
               \expandafter\xdef\csname EQ#1\endcsname
                      {\the\EQUATIONNUMBER}\fi
          Eq.~(\csname EQ#1\endcsname)\relax}
%
\newcount\SECTIONNUMBER\SECTIONNUMBER=0
\newcount\SUBSECTIONNUMBER\SUBSECTIONNUMBER=0
\def\section#1{\global\advance\SECTIONNUMBER by 1\SUBSECTIONNUMBER=0
      \bigskip\goodbreak\line{{\sectnfont \the\SECTIONNUMBER.\ #1}\hfil}
      \smallskip}
\def\subsection#1{\global\advance\SUBSECTIONNUMBER by 1
      \bigskip\goodbreak\line{{\sectnfont
         \the\SECTIONNUMBER.\the\SUBSECTIONNUMBER.\ #1}\hfil}
      \smallskip}
%
%
\begingroup\titlefont\obeylines
\hfil Large {\ititlefon q} expansion \hfill
\hfil of the 2D {\ititlefon q}-states Potts model \hfil
\endgroup\bigskip

\medskip
\centerline {\bf T.~Bhattacharya~$^a$, R.~Lacaze~$^{b,c}$, A.~Morel~$^b$ }

\vskip .5 truecm

\noindent a. MS B285, Group T-8, Los Alamos National Laboratory, NM 87544, USA

\noindent b. SPhT, CEA-Saclay, F-91191 Gif-sur-Yvette Cedex, France

\noindent c. ASCI, Bat. 506, Universit\'e Paris Sud, 91405 Orsay Cedex, France

%
\bigskip\bigskip\centerline{{\sectnfont Abstract}}\medskip
We present a recursive method to calculate a large $q$ expansion of the 2d
$q$-states Potts model free energies based on the Fortuin-Kasteleyn
representation of the model. With this procedure, we compute directly the
ordered phase partition function up to order 10 in $1/\sqrt{q}$. The energy
cumulants at the transition can be obtained with suitable resummation and
come out large for $q \lesssim 15$. As a consequence, expansions of the
free energies around the transition temperature are useless for not large enough
values of $q$.  In particular the pure phase specific heats are predicted to
be much larger, at $q \lesssim 10$, than the values extracted from current
finite size scaling analysis of extrema, whereas they agree very well with recent values
extracted at the transition point.

\bigskip\centerline{{\sectnfont R\'esum\'e}}\medskip
Une m\'ethode r\'ecursive pour calculer un d\'eveloppement \`a grand $q$ du mod\`ele de
Potts bi-dimensionnel \`a $q$ \'etats est pr\'esent\'ee, sur la base de la
repr\'esentation de Fortuin-Kasteleyn. Avec cette proc\'edure la fonction de partition
dans la phase ordonn\'ee est calcul\'ee directement \`a l'ordre 10 en $1/\sqrt{q}$.
Pour $q \lesssim 15$, les cumulants de l'\'energie sont trop importants  pour rendre
utilisable le d\'eveloppement de l'\'energie libre au voisinage du point de transition.
En particulier les chaleurs sp\'ecifiques pr\'edites pour les phases pures sont beaucoup
plus grandes que les valeurs extraites des analyses de taille finie d'extrema pour
$q \lesssim 10$, alors qu'elles sont en tr\`es bon accord avec celles  r\'ecemment
obtenues au point de transition.
\bigskip\banner

{\smallfont
 }

\noindent Submitted for publication to {\sl J. Phys. I (France)}
\vfil\eject
\section{Introduction}
The two-dimensional $q$-states Potts model is a very useful framework
for probing numerical algorithms and methods to analyze transitions.
The advantage of this model is twofold. First many of its properties, such as
the location of its transition temperature $\beta_t^{-1}$ and the internal
energy densities in the first order transition regime, are known \ref{wu}
and put strong constraints on the analysis of any numerical results.
Next, its simplicity makes it easy to implement and the existence  of the
free  parameter $q$ (the number of states) allows one to vary the properties
of the model, especially the magnitude of the correlation
length relative to accessible lattice sizes.

Two recent analytical results are of interest to our purpose.
It has been shown \ref{boko} that close to $\beta_{\rm t}$, the 
partition function $Z$ of the Potts model, in a box of volume $V=L^2$ with
periodic boundary conditions, is equal to the sum of the `partition
functions' $Z_i=\exp(V F_i)$ of the $q$ ordered and of the
disordered pure phases, up to a correction
that falls off exponentially faster with $L$. The $i^{\rm th}$ phase free
energy $F_i$ is $V$ independent and differentiable  many times with
respect to the inverse temperature $\beta$ at $\beta_{\rm t}$.
As a consequence of a recent calculation of the disordered
correlation length at $\beta_t$\ref{xi}, the interface tensions
also have been analytically predicted \ref{bojan}.

On the numerical side, various simulations have recently been performed
for different values of $q$ between 7 and 20, with the main purpose of
accumulating more experience on the identification, by numerical means, of
the nature and properties of a phase transition. Although the general overview
acquired seems quite consistent, there remains some unsatisfactory issues
such as slight inconsistencies in finite size scaling analysis
of the energy cumulants close to the transition temperature $\beta_{\rm t}^{-1}$,
and some discrepancies between exact results and numerical simulations for the
interface tension.
A  proper understanding  of these differences   is important
since in other cases of physical interest such as the 3D $q=3$ Potts
model  or  QCD at the   deconfinement transition, no good analytic
solution exists  and   one   has   to resort  to    similar  numerical
calculations to determine these quantities.

The problem at hand can be stated as that of disentangling finite size effects
coming from {\sl adding up} the ``asymptotic'' partition function $Z_i$, and
those associated with {\sl truly non-asymptotic} contributions coming from interfaces
between coexisting phases. One way towards the solution of this problem is to
learn more about the pure phase free energies $F_i(\beta )$ in the vicinity of
$\beta_t$ and as a function of $q$.

This is our motivation for starting a large $q$ expansion of $F_i$ for $\beta$
near $\beta_t$ based on the Fortuin-Kasteleyn representation \ref{kast}.
Such an expansion is much in the spirit of the pioneering work by Ginsparg,
Goldschmidt and Zuber \ref{guinsp}, who pointed out that in d-dimension, at
any finite order in $z=q^{-1/d}$, only a finite number of terms contribute in the
{\mamfont Z}$_q$ character expansion of the partition function. Low
temperature expansions also exist for the same model \ref{enting}, but
the above method seems more adapted to our goal of studying the model
close to the transition temperature. A short account of our work has been
already published with the free energy expanded in $z=1/\sqrt{q}$ up to order
9 \ref{nous} or 10 \ref{dallas}. The present paper provides a detailed
description of the expansion and an analysis of the behavior in $q$ of the
six first energy cumulants. More phenomenological consequences on finite size
analysis of numerical data, sketched in \refand{nous}{dallas}, are fully
developed in a separate publication \ref{free}.

The organization of the  paper is as follows.
In section 2 we recall some useful properties of the 2-d $q$-states Potts model
and its representation as a model of bonds. Section 3 is devoted to the detailed
computation of the partition function. The resulting free energy expansion is
given in section 4. Resummation techniques based on Pad\'e approximants are
applied in section 5 to the energy cumulants at $\beta_t$. Strong evidence is given
that they increase very fast as $q$ is lowered toward 4, the more so their order
increases. Quantitative predictions are given down to $q=6$, some of them
being in clear disagreement with values extracted from numerical simulations
not analyzed at $\beta_t$.
Concluding remarks are presented in section 6. 
 
\section{The model and its Fortuin-Kasteleyn representation.}
We consider the two-dimensional $q$-state Potts model \ref{potts}
on a square lattice with $V=L^2$ sites, defined by the partition function
$$Z=\sum_{\{\sigma_i\}}\exp{(-\beta H)},\qquad\qquad\qquad
H=-\sum_{\langle ij \rangle}\delta_{\sigma_i\sigma_j}, \EQNO{defz}$$
where $i$ and $j$ denote the lattice sites,
$\langle ij \rangle$ the pairs of
nearest neighbors and $\sigma_i=1,2,\cdots,q$.
The symmetry group of the Hamiltonian is the
permutation group of $q$ elements.

There is abundant literature on this model \ref{wu},
and many of its properties are known exactly.
In particular, it possesses a phase transition
which, for $q > 4$, is first order \ref{baxt} and lies at a coupling
$$\beta_{\rm t}=\log(\sqrt q+1).  \EQNO{defbc} $$
At $\beta=\beta_{\rm t}$, $q+1$ phases can coexist. One is
the disordered phase, the
$q$ other ones are degenerate ordered phases.
The internal energy densities at $\beta=\beta_{\rm t}$ in
the disordered and ordered phases are
$E_d$ and $E_o$ respectively, with
$$E_d+E_o=-2(1+{1\over{\sqrt q}}) \EQNO{EpE}$$
and a latent heat
$${\cal L} \equiv E_d-E_o
  =2(1+{1\over\sqrt q})\tanh{\theta\over2}\prod_{n>0}[\tanh{(n\theta)}]^2
\EQNO{defheat} $$
 where $2\cosh{\theta}=\sqrt{q}$.
Duality relates the ordered and disordered free energies
$$F_d(\tilde\beta)=F_o(\beta)-2\ln\bigl(({\rm e}^{\beta}-1)/\sqrt{q}\bigr)
\qquad,\qquad ({\rm e}^{\tilde\beta}-1) ({\rm e}^{\beta}-1)=q.\EQNO{dual}$$

A large $q$ expansion can be obtained through the 
Fortuin-Kasteleyn \ref{kast} representation of the Potts model partition
function
$$Z=\sum_{X} ({\rm e}^{\beta}-1)^{b(X)} q^{c(X)} \EQNO{FK}$$
where $X$ is any configuration of bonds on a square
lattice, $b(X)$ is the number of bonds in configuration $X$, and $c(X)$ its
 number of 
clusters of sites : {\sl two sites bound to each other belong to the same
cluster} ( an isolated site is a cluster).

The completely ordered configuration, $X_{\rm  o}$, has all possible bonds
and, hence, corresponds to $c(X_{\rm o})=1$ and $b(X_{\rm o})=2 V$.
So the partition function of the ordered phase, $Z_{\rm o}$
can be reorganized as an  expansion in $q^{-{1\over2}}$ about $X_{\rm o}$:
$$Z_{\rm o} = q ({\rm e}^{\beta}-1)^{2V} \sum_{k\geq 0,l\geq 0} N^{\rm o}_{k,l}(V)
    ({{\rm e}^{\beta}-1\over\sqrt{q}})^{-l}q^{k-{l\over2}}, \EQNO{Zo}$$
where $N^{\rm o}_{k,l}(V)$ is the number of configurations in a volume $V$ with $l$
bonds removed and comprising $k+1$ clusters.
We made explicit in  \eq{Zo} the factor $({\rm e}^{\beta}-1)/\sqrt{q}$ which
is 1 at the transition and thus provides an expansion of $Z_{\rm o}$
in $q^{-1/2}$ near the transition.
The enumeration of all the $N^{\rm o}_{k,l}(V)$  such that
$(l-2k)\leq M$ yields this expansion to order $M$ in $q^{-1/2}$.

 Similarly, the completely disordered configuration, $X_{\rm  d}$,
corresponds to $c= V$ and $b=0$.
Hence we reorganize the expansion of the partition function of the
disordered phase about this configuration at temperature
$\tilde\beta^{-1}$ as
$$Z_{\rm d} = q ^{V} \sum_{j\geq 0,l\geq 0} N^{\rm d}_{j,l}(V)
    ({{\rm e}^{\tilde\beta}-1\over\sqrt{q}})^{l}q^{{l\over2}-j}, \EQNO{Zda}$$
where $N^{\rm d}_{j,l}(V)$ is the number of configurations in a volume $V$ with $l$
bonds and $V-j$ clusters. Since the suppression of $j$ clusters requires that
{\sl at least} $j$ bonds are restored, one has $k\equiv l-j\geq 0$.
Hence we rewrite $Z_{\rm d}$ as
$$Z_{\rm d} = q ^{V} \sum_{l\geq 0,k\geq 0} \tilde N^{\rm d}_{k,l}(V)
    ({{\rm e}^{\tilde\beta}-1\over\sqrt{q}})^{l}q^{k-{l\over2}}, \EQNO{Zd}$$
where $\tilde N^{\rm d}_{k,l}(V)$ is the number of configurations in a volume $V$ with $l$
bonds and $V-l+k$ clusters.
Duality as given in \eq{dual} implies $\tilde N^{\rm d}_{k,l}(V)=N^{\rm o}_{k,l}(V)$,
a trivial geometrical property exemplified later on. Thus, in the following
we restrict our calculation to the ordered phase contributions,
except for comments on duality.

With a large enough volume $V$ (to eliminate all finite size effects)
and periodic boundary conditions (to eliminate edge effects),
$N(V)$ are polynomials in $V$ to any given finite order $M$,
and all configurations retained  correspond to {\it disordered (ordered)
islands} in a bulk {\it ordered (disordered) phase}.

\section{Evaluation of the expansion}

This simple expansion which computes directly $Z_{\rm o}$ can be made
recursive. To this end we explicitly remove successively up to 8 bonds
in sub-section 3.1 where the main properties are exhibited in order to make
the general construction in sub-sections 3.2-3.6. To simplify,
the number $k$ of clusters added to the bulk ordered cluster will be called
in the following the number of cluster.

\subsection{Explicit first steps}
Starting from the ordered configuration, we have $2V$ ways to remove one bond,
the number of clusters remaining 0, so that $N^{\rm o}_{0,1}(V)=2V$.
Continuing to remove bonds, we obtain $N^{\rm o}_{0,2}(V)=2V (2V-1)/2$ and
$N^{\rm o}_{0,3}(V)=2V (2V-1) (2V-2)/6$. To proceed, let us introduce
$C(n,l)$ as the number of ways to remove $l$ identical bonds
among $2V-n$ ones
$$
C(n,l)={1\over l!} \prod_{i=0}^{l-1} (2 V-n-i), \EQNO{defC}$$
\line{\hfil\vbox{
\vbox{\hsize=8.5 truecm\parfillskip=0pt
Thus
$$\eqalign{
N^{\rm o}_{0,1}(V)=&C(0,1),\cr
N^{\rm o}_{0,2}(V)=&C(0,2),\cr
N^{\rm o}_{0,3}(V)=&C(0,3). }$$

A new situation occurs with 4 bonds removed.
We have $C(0,4)$ ways to remove them, but some of them give rise to
an isolated 1-site cluster as shown in \fig{1.04}, when the 4
}}
\hfill\vbox{
\vbox{\hsize=5.0 truecm\parfillskip=0pt
\centerline {\bf \fig{1.04}}
\noindent {\ismallfont Four removed bonds (crosses) making a 1-site cluster.\hfill}
}}
\hfil}
\includegraphics{1.04.ps}
\vskip -0.60cm 
\noindent
removed bonds have one site in common. The number of ways to make
this figure is V, the number of possible positions for the one site cluster.
Thus we have with 4 removed bonds
$$\eqalign{
N^{\rm o}_{1,4}(V)=&V, \cr
N^{\rm o}_{0,4}(V)=&C(0,4)-V, }$$
where the number of configurations which do not correspond
to the specified number of clusters in $N^{\rm o}_{0,4}$ is subtracted.

With 5 removed bonds, nothing new happens and one has
$$\eqalign{
N^{\rm o}_{1,5}(V)=&V C(4,1), \cr
N^{\rm o}_{0,5}(V)=&C(0,5)-V C(4,1), }$$
\noindent where the factor $C(4,1)$ takes into account the number of ways to
remove one bond when 4 bonds have already been removed to make the configuration
of \fig{1.04}.
\smallskip
\line{\hfil\vbox{
\vbox{\hsize=8.5 truecm\parfillskip=0pt
 With 6 removed bonds, the new type of configuration shown in
\fig{1.06} can be made with $2V$ possibilities ($V$ translations and 2 orientations),
and its corresponds to one cluster (the 2 sites surrounded by the removed bonds
are in the same cluster).
{Thus we have \hfill}
$$\eqalign{
N^{\rm o}_{1,6}(V)=&V C(4,2) +2 V ,\cr
N^{\rm o}_{0,6}(V)=&C(0,6)-V C(4,2) -2 V .}$$
}}
\hfill\vbox{
\vbox{\hsize=5.0 truecm
\centerline {\bf \fig{1.06}}
\noindent {\ismallfont Six removed bonds (crosses) making a 2-site cluster.}
\vskip 0.1cm}
}
\hfil}
\includegraphics{1.06.ps}
\noindent Here again the multiplicity $C(0,6)$ is corrected for
all the 1 cluster contributions.
\smallskip
\line{\hfil\vbox{
\vbox{\hsize=8.5 truecm\parfillskip=0pt
 Next with 7 bonds removed, the higher contribution comes from the configuration
of \fig{2.07} with 2 clusters and with $2V$ possibilities, and thus $N^{\rm o}_{2,7}=2V$.
The contributions to $N^{\rm o}_{1,7}$ are obtained from the configurations contributing
to $N^{\rm o}_{1,6}$ with one more bond removed. First we have V C(4,3) configurations made of \fig{1.04} with
3 more  bonds removed, out of which we have to
subtract the $4V$ configurations with 2 clusters
as in \fig{2.07} (the 3
}} 
\hfill\vbox{
\vbox{\hsize=5.0 truecm 
\centerline {\bf \fig{2.07}} 
\noindent {\ismallfont Seven removed bonds (crosses) making two 1-site clusters.} 
} }
\hfil} 
\includegraphics{2.07.ps}
\vskip -0.65cm 
\noindent 
removed bonds can surround the 4 neighboring
sites of the 1-site cluster of \fig{1.04}and these configurations do not
give new contribution to $N^{\rm o}_{2,7}$ for which the correct counting
is already made).
Next we have to remove one bond to the $2V$ configurations of \fig{1.06}.
If we remove the ``internal'' bond connecting the two sites of the cluster,
we will get 2 clusters. So we only remove bonds of the bulk ordered phase,
that is only $2V-7$ bonds can be removed.
Thus we have 
\smallskip
\line{\hfil\vbox{
\vbox{\hsize=8.5 truecm\parfillskip=0pt
$$\eqalign{
N^{\rm o}_{2,7}(V)=&2 V ,\cr 
N^{\rm o}_{1,7}(V)=&V C(4,3) -4 V +2 V C(7,1) ,\cr 
N^{\rm o}_{0,7}(V)=&C(0,7)-N^{\rm o}_{1,7}(V)-N^{\rm o}_{2,7}(V) .}$$ 

As a last simple example, let us consider 8 removed bonds and discuss the new
contributions beside the contributions deduced from
$N^{\rm o}_{k,7}$ with the appropriate changes in the $C$
}}
\hfill\vbox{
\vbox{\hsize=5.0 truecm 
\centerline {\bf \fig{i1.04}} 
\noindent {\ismallfont Sites forbidden (circles) for a second 1-site cluster.}
\vskip 0.1cm} 
}
\hfil} 
\includegraphics{i1.04.ps}
\vskip -0.65cm 
\noindent 
factors.
New configurations with 2 clusters can be made by putting the \fig{1.04} twice
on the lattice. Once the first figure is put on the lattice ($V$ possibilities),
the second one
cannot be
put at the same site nor at the 4 neighboring sites as shown by open dot in
\fig{i1.04} because a bond ending there has already been removed. Thus the number
of such configurations is $V (V-5)/2$ with the 1/2 factor for symmetry.
Concerning the contribution $N^{\rm o}_{1,8}$, we first correct the
contribution coming from \fig{1.04} plus 4 removed bonds forming themselves
a 1-site cluster, that is a correction $V (V-5)$ ( no 1/2 symmetry factor).
Then we add the contribution coming from the configurations shown in 
\fig{1.08} where the cluster extends either over 3 sites at left ( $6V$
 possibilities) or over 4 sites at right ($V$ possibilities). Thus
$$\eqalign{
N^{\rm o}_{2,8}(V)=&2 V C(7,1)+{1\over 2}V (V-5),\cr
N^{\rm o}_{1,8}(V)=&V C(4,4) -4 V C(7,1) -V (V-5) +2 V C(7,2) +6 V +V,\cr
N^{\rm o}_{0,8}(V)=&C(0,8)-N^{\rm o}_{1,8}-N^{\rm o}_{2,8}. }$$
In $N^{\rm o}_{1,8}$ we keep the two contributions of \fig{1.08} separate because
at the next step, as we explained for $N^{\rm o}_{1,8}$ from \fig{1.06} with one
more removed bond, it is easiest
to restrict the removed bond to belong to the bulk ordered phase, that is
$2 V -10$ and $2 V -12$ possibilities respectively.
\vskip 5.5 truecm
\includegraphics{1.08f3.10.ps}
\includegraphics{1.08f4.12.ps}
\centerline{\bf \fig{1.08}}
\centerline{\ismallfont One cluster with 8 removed bonds, extending}
\centerline{\ismallfont over 3 sites at left and 4 sites at right.}
\smallskip

These examples illustrate the way how a recursive construction of the expansion
up to a given order $M$ can be made. This construction involves 5 steps:
\item{i)} Construct the {\bf dominant configurations} made of one connected set
of 1-site clusters (two clusters with neighboring sites belong to the same
connected set ). We note ($k$,$l$) the set of configurations with $k$ connected
 1-site clusters and $l$ removed bonds and they are constructed for all the $k$
 and $l$ such that $l-2 k\leq M$.
Thus \fig{1.04} and \fig{2.07} belong respectively to (1,4) and (2,7).
\item{ii)} Construct the {\bf sub-dominant configurations} made of one connected set
of clusters involving at least one cluster with more than one site
( cf. \figand{1.06}{1.08} ). They can be obtained by restoring one or several
internal bonds of dominant (parent) configurations and we note ($k,l;k^p,l^p$)
the configuration set with $k$ clusters and $l\leq M-2k$ removed bonds inside
 a parent with $k^p$ clusters and $l^p$ removed bonds. Thus the configuration of
\fig{1.06} belongs to (1,6;2,7), those in \fig{1.08} left and right to
(1,8;3,10) and (1,8;4,12) respectively.
\item{iii)} Construct the {\bf product configurations} which are disconnected sets of clusters. Once we have 
all the connected sets, we compute the  the number of ways to put together on the lattice
connected sets of clusters with $l_i$ and $k_i$ such that $\sum l_i-2\sum k_i\leq M$.
For example the contribution $V(V-5)/2$ to $N_{2,8}$ found above is the 
product $(1,4)*(1,4)$.
\item{iv)} Construct the {\bf correcting contributions} which are obtained from the
configurations with $k$ and $l$ of steps i) to iii) plus $n$ removed bonds
($1\leq n\leq M+2k-l$) and have a number of clusters higher than $k$.
These correcting contributions will allow to get from
the contributions of configurations obtained in steps i) to ii) weighted by
$C(l,n)$ the correct contribution to $N^{\rm o}_{k,l+n}$ by suppressing
the contributions from higher number of clusters (as the correction $4V$ in the
contribution to $N^{\rm o}_{1,7}$ of \fig{1.04} plus 3 removed bonds).
\item{v)} {\bf Collect the results} of steps i) to iv) to get all the $N_{k,l}$'s
relevant to order $M$. In the steps i) to iv) the number of removed bonds is
minimal in the sense that they are all necessary to make the clusters.
Thus this step collects the preceding contributions with appropriate signs
(for correction) and $C$' factors.

These different steps are explained in details in the following sub-sections.
Of course any configuration drawn on the lattice has V copies obtained by the
lattice translations and two configurations will be said distinct if they are
unequivalent upon lattice translation.

\subsection{The dominant configurations}
If we consider all the connected sets of clusters corresponding to
$k$ clusters with $l$ removed bonds and extending on $n$ given 
(connected) sites, the exponent $l-2k$ of $q_{-1/2}$ in $Z_{\rm o}$ is
minimum for $k=n$. Hence the name of dominant for a ($k$,$l$) set of 
configuration extending on $k$ sites and thus made of $k$ 1-site clusters.
All the dominant ($k$)--configurations as defined by the data of $k$ connected
sites can be obtained recursively from the ($k-1$)--ones
\item{i)}connecting one additional site in all
the possible ways to all the ($k-1$) configurations, keeping only the distinct (upon translation) configurations.
\item{ii)}eliminating those configurations which happen to have more than $k$
clusters ( extra sites can be isolated from the bulk ordered ones and these
configurations can
be identified by a suitable cluster finding algorithm \ref{kopel} ).

In this construction, a dominant $(k,l)$--configuration contributing at order
$m=l-2k$ can be obtained from one or
several of the following dominant $k-1$--configurations 

\item{  i)}  $(k-1,l-3)$ ~~~contributing to order $m-1$
\item{ ii)}  $(k-1,l-2)$ ~~~contributing to order $m$
\item{iii)}  $(k-1,l-1)$ ~~~contributing to order $m+1$

\noindent An example of case i) is given by \fig{2.07} obtained from \fig{1.04},
while examples of case ii) and iii) are given in \fig{6.17} left and right
respectively.
\vskip 5.5 truecm
\includegraphics{6.17f5.15.ps} 
\includegraphics{6.17f5.16.ps} 
\centerline{\bf \fig{6.17}}
\noindent{\ismallfont The same (6,17) dominant contribution obtained by connecting
a $6^{th}$ 1--site cluster (open point) to a (5,15) (left, case ii) or to a
(5,16) (right, case iii).}
\smallskip
Note that in the latter case, the order in $q^{-1/2}$ decreases in the step
$k-1 -> k$. Hopefully,
the example of \fig{6.17} is in fact generic, and
one can convince oneself that a $(k,l)$--configuration obtained via iii) 
can always be obtained via either ii) or  i).
This means that, once the maximum order $M$ of the
computation is given, it is never necessary to keep contributions to a higher
order in the iterative process.

Once all the distinct dominant configurations with $k$ 1-site clusters are obtained,
they can easily be classified according to the number $l$ of removed bonds.
Thus a $(k,l)$ set is given by a set of $2k$ data, the $x$ and $y$
positions of the $k$ 1-site clusters.
Their number is
given in Table~1 and represents the corresponding contribution to
$N^{\rm o}_{k,l}/V$. At order $M=10$, 59 $(k,l)$ set are contributing, the
largest size to be considered being a (25,50) configuration, a $5\times 5$
square of 1--site clusters.

\subsection{The sub-dominant configurations}
Sub-dominant configurations correspond to connected set of clusters
involving at least one cluster extending over more than one site. 
They can be obtained from the dominant
configurations by restoring some of their $4k-l$ internal links.

Let us start from a $(k^p,l^p)$ dominant set, the parent of a family
of sub-dominant $(k,l;k^p,l^p)$ configurations. If one interior bond
is restored, the two sites it links now belong to the same (2-sites)
cluster. Then $k=k^p-1$ and $l=l^p-1$ and the exponent is one unit
less that the parent one ( $m=l-2k=m^p-1$ ). Hence the name subdominant.
For example \fig{1.06} can be obtained from \fig{2.07} by restoring the
interior bond (only one possibility),leading to a  2-site cluster.
 
These sub-dominant configurations are obtained from the dominant
$(k^p,l^p)$ configurations by restoring the interior bonds one
after the other in all possible distinct ways, looking at the $k$
and $l$ values of the generated configurations.
In fact, it is not necessary to keep memory of the
restored bonds and for each $(k^p,l^p)$ configuration it is enough to count
the number of distinct $(k,l;k^p,l^p)$ configurations obtained.
Thus a $(k,l;k^p,l^p)$ set is given by a set of $2k^p+1$
data, the $x$ and $y$ positions of the parent $k^p$ 1--site clusters and
their multiplicity.

Restoring systematically bonds among the $4k^p-l^p$ interior ones can be
time consuming for large clusters. However restoring up to 5 bonds can
be implemented  easily because there is a classification according
to the number of 4 1-site cluster making a plaquette, with the property
that all the parent configuration give rise to the same multiplicity.

For the order $M=10$ considered here, we had to construct 146 sets
($k,l;k^p,l^p$) from the 59 dominant ($k,l$).

\subsection{The product configurations} 
 
Let us now consider disconnected configurations. They are made of components
which are either dominant or subdominant configurations. In fact for
components which are sub-dominant configuration, the corresponding result is the same as given by the 'parent' component, and thus we can only consider
product of dominant contributions. Here ``disconnected'' implies that two requirements must be fulfilled
\item{i)} no overlap between links of any two of the connected component;
\item{ii)} no site of the surrounding bulk ordered phase become a cluster.

According to these requirements, the values of $k$ and $l$ of a product are
the sums of the $k_i$'s and $l_i$'s of the factors (its components). The method used is first to count all the possible ways the components can be put on the
lattice with requirement i) satisfied, then subtract the number of configurations which do not satisfy $k=\sum k_i$.

For the product of two configurations given by the sets $F_1$ and $F_2$ of
occupied sites, once the first configuration $F_1$
is put on the lattice with $V$ possibilities, we have to search all the
possibilities to put the second one. For that purpose let us define as $F_{12}$ the
set of sites around $F_1$ where a given site of $F_2$ taken as origin cannot be
put according to requirement i). If $v_{12}$ is the number of sites in $F_{12}$, then
we have $V (V-v_{12})$ configurations for the product $F_1\star F_2$ (divide by 2
for identical configurations).
An example is the
product (1,4)$\star$(1,4) as already been given in subsection 3.1.

This process can be generalized to product of more figures.
For the product of $F_1\star F_2\star F_3$ let us first define the protected areas
$F_{12}$, $F_{13}$ and $F_{23}$ with $F_{ij}$ the set of $v_{ij}$ sites around $F_i$
which cannot be occupied by an origin site of $F_j$ according to requirement i).
Once $F_1$ is on the lattice ($V$ possibilities) let us first put $F_2$ far enough
from $F_1$ such that $F_3$ can be put every where except on $F_{13}$ or $F_{23}$.
This will be the case when $F_2$ is every where except in a region $F_{12,23}$
for which there will be overlapp of sites of $F_{13}$ and $F_{23}$. If $v_{12,23}$
is the number of sites of $F_{12,23}$, we get a first number of possible
configurations which is 
$$n_a=V (V-v_{12,23}) (V-v_{13}-v_{23}).$$
Next we consider all the possible positions of $F_2$ inside $F_{12,23}$ but outside
$F_{12}$ ($v_{13,23}-v_{12}$ possible positions). For each such position there is
a protecting region $F_{(12)3}$ with $v_{(12)3}$ sites forbidden for $F_3$
according to rule i) ($v_{(12)3}<v_{13}+v_{23}$) and we have to add to $n_a$
$$n_b=V \sum_{i=1}^{v_{13,23}-v_{12}} (V-v_{(12)3}^i).$$

This method can be extended to products of more factor and implemented on computer.
As an example the resulting polynomials in $V$ for the powers of (1,4) needed
at order $M=10$ are
$$\eqalign{
(1,4)^2=&{1\over 2}V(V-5),\cr
(1,4)^3=&{1\over 6}V(V^2-15V+62),\cr
(1,4)^4=&{1\over 24}V(V^3-30V^2+323V-1254),\cr
(1,4)^5=&{1\over 120}V(V^4-50V^3+995V^2-9370V+35424).\cr
  }\EQNO{prod1.4}
$$

Finally we have to fulfill requirement ii), and thus subtract to the above
configurations those which
do not correspond to the specified number of clusters, $\sum k_i$.
Two examples are given in \fig{nprod}, with at left a peculiar configuration
of $(1,4)^4$ contributing to a 5 clusters contribution and at right
a product $(3,10)^2$ contributing to 7 clusters.
\vskip 6.5 truecm
\includegraphics{1.04x4.ps}
\includegraphics{3.10x2.ps}
\centerline{\bf \fig{nprod}}
\centerline{\ismallfont Product of configurations giving rise to extra cluster number.}

\noindent These corrections can be obtained by trying to cast dominant or
subdominant configurations with $l=\sum l_i$ into the corresponding products
(see next subsection).
\smallskip

\subsection{Correcting configurations} 
In the three preceding sub-sections, it was assumed that the number
of removed bonds was minimum to make a definite configuration of clusters.
In other words there is no ``free'' links, that is removed bonds not
belonging to any cluster.
Thus the number $n_{(k,l)}$ of configurations ($k,l$) gives directly
its contribution to $N_{k,l}$
$$ \bigl[{\rm contrib.}\ (k,l)\ {\rm to}\ N_{k,l}\bigr] = n_{(k,l)}.$$
(For simplicity we consider only dominant configurations).
We now wish to get its contribution to $N_{k,l+i}, 1\leq i\leq M+2k-l$.
This contributions is $n_{(k,l)} C(l,i)$  up to corrections from
the configurations corresponding to a higher value of cluster number.
These corrections are twofold. First these $i$ removed bonds can make a cluster
and the corresponding corrections can be computed as in the preceding subsection.
Or the configuration $(k,l)$ with $i$ removed bonds can make connected sets of
clusters which can be considered as inside dominant configurations. Thus in \fig{2.7+5}
a (2,7) configurations with 5 more removed bonds can make connected sets of clusters
inside a (4,12) at right or inside a (4,13) at left. The distinction between
the different possible parent dominant configurations is necessary for further
removing of bonds.
\vskip 6.0 truecm
\includegraphics{2.07+5a.ps}
\includegraphics{2.07+5b.ps}
\centerline{\bf \fig{2.7+5}}
\centerline{\ismallfont (2,7) configuration plus 5 removed bonds making connected}
\centerline{\ismallfont sets of clusters inside a (4,12) at right or a (4,13) at left.}

If $n_{((k,l)+i;k^p_r,l^p_r)}$ is the number of these $r$ configurations , then
$$ \bigl[{\rm contrib.}\ (k,l)\ {\rm to}\ N_{k,l+i}\bigr] = n_{(k,l)}C(l,i)-n_{((k,l)+i;k^p_r,l^p_r)},$$
with eventual summation on $r$.
This process has to be continued and let us suppose that the configuration
${((k,l)+i;k^p_r,l^p_r)}$ plus $j$ removed bonds can make higher configurations
$({((k,l)+i;k^p_r,l^p_r)}+j;k^p_s,l^p_s)$. Then there is also higher configurations
obtained directly from $(k,l)$ plus $i+j$ removed bonds inside the same parents.
The proper correction which take correctly the $i+j$ bonds symmetrization is the
direct one. This means that we have to correct the correction itself on the form
$$\eqalign{
\bigl[ {\rm contrib.}\ (k,l)\ {\rm to}\ &N_{k,l+i+j}\bigr] =  n_{(k,l)}C(l,i+j)\cr
-&\bigl[(n_{((k,l)+i;k^p_r,l^p_r)}C(l^p_r,j)-n_{(((k,l)+i;k^p_r,l^p_r)+j;k^p_s,l^p_s)})\bigr]\cr
-&n_{((k,l)+i+j;k^p_s,l^p_s)}.\cr
}$$
with eventual summation on $r$ and $s$.

Clearly this computation apply to dominant or sub-dominant contribution as
well as to product of connected sets of clusters and
for all the configurations from subsections 3.2 to 3.4 characterized by
$k$ and $l$, we have in this step to compute recursively all the connected
sets of clusters obtained with $l_f$ more removed links ($1\leq l_f\leq M+2k-l$).
The recursive process means that we start from all the already obtained
configurations with $(k,l)$ plus $l'_f$ removed bonds ($0\leq l'_f<l_f$) and add $l_f-l'_f$ removed bonds.
 
Such configurations
will be in the list of dominant or sub-dominant contributions. So as
in section 3.3, we have to search all the possibilities to found a starting
configuration plus the removed links inside the dominant configurations in such
a way that all the exterior links match. In the example of a (2,7) configuration and
5 removed bonds, we know from the sub-dominant configuration analysis that a connected set
of clusters with 12 removed bonds and more than 2 clusters can be inside a (4,12) or a
(4,13). For the (4,12) there is 2 possibilities to found the (2,7) configuration of
\fig{2.7}. For the (4,13) configurations, we have to search the possibilities to put
\fig{2.7} inside the (4,13) with one restored interior link and found that
3 (4,13) configurations give
 rise to 2 possibilities and 8 other ones to 1 possibility.

\subsection{Collecting the contributions to $N_{k,l}$.}
In computing the partition function from \eq{Zo}, we need the value of $N_{k,l}$
for each values of $k$ and $l$. The contributions to $N_{k,l}$ comes from the
dominant, sub-dominant and product configurations obtained in subsections 3.2-3.4
and corresponding to a  number $k_c$ of clusters and  a number $l_c$ of removed
bonds such that $k_c=k \ ,\ l_c\le l$, up to corrections as explained in subsection 3.5.
The whole procedure is better understood on a particular example and let us consider
the computation of the $N_{2,12}$ coefficient. Its contributions comes from dominant,
sub-dominant and product configurations with $k=2$ and $l\leq 12$ which can be
\item {.} (2,7) dominant configurations,
\item {.} {2,9), (2,10), (2,11) and (2,12) sub-dominant configurations,
\item {.} (1,4)$\star$(1,4), (1,4)$\star$(1,6), (1,4)$\star$(1,8) and (1,6)$\star$(1,6) product configurations.

\noindent For each of them, we have to correct their naive weight (number of configurations
times the appropriate $C$ factor of Eq. \eq{defC} for the extra links)
for the configurations corresponding to a higher
number of clusters. The various contributions building up the final value of $N_{2,12}/V$
are given in the first column of Table~2, along with its corresponding set of cluster in
column 2 (the last column labels the lines). Note that when a $C$ factor is present, its
second argument gives the number of free links, that is the number of removed bonds
not belonging to a cluster.

The first contribution (line 1) comes from the dominant (2,7) configurations (two such
configurations weighted by $C(7,5)$, the number of ways to removed 5 bonds among $2V-7$ ones).
Its corresponding corrections (contributions
corresponding to a number of cluster higher than 2) are given in the lines 2 to 7.
First in line 2, 4 removed bonds can isolate a 1-site cluster as in \fig{1.04}.
For each (2,7) configuration there is $(V-8)$ such possibilities ( the product
$(2,7)\star (1,4)$ ) with 1 link remaining free among $2V-11$.
Lines 3-6 concern clusters obtained with (2,7) and 3 to 5 removed bonds. We made
a distinction between different possible parents in lines 5 and 6 because they
imply different $C$ factor for their corresponding contributions at $N_{2,13}$.
Furthermore these (2,7)+5 corrections represent the whole contribution to 
higher cluster numbers obtained with (2,7) and 5 extra removed bonds such that
all the 5 extra links belong to the clusters. However in the line 3
such corrections are included and should not have been subtracted. Line 7
provides this ``correction to the correction'' of line 3, and thus with a
positive sign.

Lines 8 and 11 to 15 concerns contributions of sub-dominant contributions.
They are characterizes by their parent distribution in order to determine
the number of frozen links which give the arguments of the $C$ coefficients.
Only line 8 needs subtraction in line 9 and 10, either 2 or 3 extra links giving 
rise to $k>2$ configurations.
Let us note that there is two ways to have a 2 clusters inside a dominant (5,15),
one in line 14 when restoring 3 bonds, one in line 13 when restoring the four bonds
of a loop.

We finally start with The possible products of dominant or sub-dominant $k=1$
configurations ( lines 16,23,25-27). First in the product (1,4)$\star$(1,4)
we consider the two (1,4) as different in such a way that all the corresponding
contributions (lines 16-22) have an explicit 1/2 factor. The first correction
concern the 4 extra removed bonds making a (1,4) cluster. This product (1,4)$^3$
have no extra symmetry factor, the third (1,4) factor being formed by the extra
links distinctly from the two other (identical) factors. The connected cluster
corrections are in line 18-21, with again a ``correction to the correction''
with a positive sign in line 21. We have also to make a disconnected cluster
correction in line 22,  when 3 of the extra links are making a 2 clusters with
each of the product configurations. For the other products only one of them
need to be corrected.

All these computations have been made automatic on workstation.
\section{Results}
 We have computed the series for $Z_o$ up to order $M=10$ in $q^{-1/2}$,
which involves contributions up to $N_{60,25}(V)$, that is up to 60 removed
bonds and 25 clusters. This series exponentiates in $V$
to give $F_o$  to order $M$ in the form
$$F_{\rm o}(\beta)\equiv {1\over V}\ln Z_o = 2\ln({\rm e}^{\beta} -1)+
\sum_{m\geq 1,l\geq 1} A_{l,m} ({{\rm e}^{\beta}-1\over\sqrt{q}})^{-l}
q^{-m/2} + {\cal O}(q^{-11/2}), \EQNO{Fo}$$
where the non-zero coefficients $A_{l,m}$ are given in Table~3 up to $m=10$.

One may notice, at the top of the columns, the appearance of stable sequences
with decreasing $l$.
For $k$ even, the $m$/2 higher $l$ values are in the sequence
1, 6, 22, 68, 187, ...  that is as long as the contributions
are coming from removing a corner (case ii) when starting from a square of size
$m$/2~$m$/2. For $m$ odd, the series is 2, 8, 30, 88, ... for the ($m$-1)/2 
higher $l$ values for the
same reason when starting from a rectangle ($m$+1)/2~($m$-1)/2. In contrast the
bottom of each column is totally alternating. 

Our series truncated at $l=20$ can be compared to the result of 
Ref. \ref{enting}. Up to order 9 in $q^{-1/2}$, the two series are in
agreement. At order 10, some coefficients, missing in \ref{enting},
cannot be compared, but we disagree 
with the coefficient of $r_3 u_{17}$ of Ref. \ref{enting} (we find 3822
instead of 3818) and of $r_2 u_{16}$ (6269 instead of 6265).

Our result is obtained for the ordered phase. Then the disordered free energy can
be obtained using the duality relation given in \eq{dual}. As already mentioned
our geometrical construction must comply with duality, and this
can be explicitly verified on an example as shown in \fig{5.15} for a dominant
configuration (5,15).
\vskip 6.5 truecm
\includegraphics{5.15.ps}
\includegraphics{d5.15.ps}
\centerline{\bf \fig{5.15}}
\centerline{\ismallfont (5,15) dual configurations in the ordered and disordered phase}
\item a) At left, 15 bonds removed from the fully ordered phase increases
the number of clusters by 5 and contributes to $N^{\rm o}_{15,5}(V)$ of \eq{Zo}.
\item b) At right, 15 bonds restored from the fully disordered phase  make
one cluster only out of 11 sites, reducing the number of clusters by 10,
and thus contributing to $N^{\rm d}_{10,15}(V)$ of \eq{Zda}, or
$\tilde N^{\rm d}_{5,15}(V)$ of \eq{Zd}. 
The open dots in the disordered figure (at right) are the centers of the
bond plaquettes. They coincide with sites of the dual lattice, reproducing the same
figure as in the left. Thus the contribution to $N^{\rm o}_{5,15}(V)$ and
$\tilde N^{\rm d}_{5,15}(V)$ of such figures are the same as a priori stated
from duality.
\bigskip
\bigskip
The expansion of the free energy gives similar series for the
$n^{\rm th}$ derivative with respect to $\beta$, $F_{\rm o}^{(n)}(\beta )$.
At $\beta=\beta_{\rm t}$ we write the energy cumulants as

$$
F_o^{(n)}\equiv F_o^{(n)}(\beta_t)=(\ln(q)+2 z)\delta_{n0}+2\delta_{n1}+(-)^n
\sum_{m=2}^{10}C_m^nz^m+{\cal O}(z^{11}), \EQNO{Fon}$$
for
$$ z={1\over \sqrt{q}}$$
\noindent where the $C_m^n$ coefficients are given in Table~4 for $n$ varying
from 0 to 6.

The corresponding disordered cumulants can be obtained from
the ordered ones by use of the duality relation \eq{dual}. The relation
for the energies ( $E_i=-F_i^{(1)}$ ) is given in \eq{EpE}. For higher
cumulants up to $n=5$ we can write
$$\eqalign{
F_d^{(2)}-F_o^{(2)}=&{-1\over q^{1/2}}\bigl[F_d^{(1)}-F_o^{(1)}\bigr]\cr
F_d^{(3)}+F_o^{(3)}=&{q^{1/2}-1\over q}\bigl[F_d^{(1)}+F_o^{(1)}\bigr]
-{3\over q^{1/2}}\bigl[F_d^{(2)}+F_o^{(2)}\bigr] \cr
F_d^{(4)}-F_o^{(4)}=&{6-q\over q^{3/2}}\bigl[F_d^{(1)}-F_o^{(1)}\bigr]
-{6\over q^{1/2}}\bigl[F_d^{(3)}-F_o^{(3)}\bigr] \cr
F_d^{(5)}+F_o^{(5)}=&{q^{3/2}+q+24q^{1/2}-24\over q^2}\bigl[F_d^{(1)}+F_o^{(1)}\bigr]
-{5(q-12)\over q^{3/2}}\bigl[F_d^{(2)}+F_o^{(2)}\bigr]\cr
&\qquad -{10\over q^{1/2}}\bigl[F_d^{(4)}+F_o^{(4)}\bigr].}
\EQNO{dumom}$$

The $n=0$ (free energy) and $n=1$ ( internal energy) series
match the exact results \ref{baxt} up to $M=10$.
The next section is devoted to a study of the series $n\ge 2$.
\section{Resummation of the large {\isectnfon q} series}

We start from the result \eq{Fon} which gives the cumulants $F_o^{(n)}$ of the
ordered free energy $F_o(\beta)$, taken at $\beta_t$, as series expansions to
order 10 in the variable $z=1/\sqrt{q}$. The coefficients $C_m^n$ of $q^{-m/2}$
in $F^{(n)}$ are given in Table~4 up to $n=6$. We want to explore the behavior
in $q$ of $F^{(n)}$, $n\geq 2$ as $q$ is decreased towards $q=4$. The first
cumulants $F_o(\beta_t)$ and $F_o^{(1)}=-E_o$ are known exactly.
For later convenience we write $F_o^{(1)}$ as
$$F_o^{(1)}=(1+z) + {1\over 2} {\cal L},
$$
where the latent heat ${\cal L}$ is given by \eq{defheat},
with $\theta$ defined through
$$ 2 \cosh \theta = \sqrt{q}.$$

At first glance to Table~4, the task of resumming the series $F^{(n)}$ ( from now
on we omit the index $o$) looks quite discouraging: not only increase all the (known) $C_m^n$'s very fast with $m$, the more so $n$ is large, but also they are all
positive for $n\geq 2$. We will undertake this task, however, with the help of a few
assumptions on the singularities in $q$ of the $F^{(n)}$'s, and after checking that
the techniques used work well for the known case of ${\cal L}$.

From \eq{defheat}, we known that the radius of convergence of the series in
$q^{-1/2}$ of ${\cal L}$ is 1/2 ($q=4$). We will {\sl assume} that it is so
for all the $F^{(n)}$'s. Furthermore, the leading singularity of ${\cal L}$
 at $q=4$ is given by
$$
{\cal L}\sim (1+z) \tanh ({\theta\over2} ) {2\pi\over \theta} x^{-1/2},  \EQNO{Lq} $$
where
$$x=\exp ( {\pi^2\over 2\theta}) \EQNO{defx}.$$

A similar singularity at $q\rightarrow 4$ occurs in the disordered phase
correlation length \refand{xi}{bojan}
$$\xi_d\sim{1\over 8\sqrt{2}}x.$$

Our second assumption will be that for $n\geq 2$, $F^{(n)}$ diverges at
$q\rightarrow 4$  like a power of $x$, up to a smooth factor. Arguments for
that have been given in \refand{nous}{free}, where we proposed that
$F^{(n)}/x^{3n/2-2}$ is a slowly varying function of $x$.
\vfill\eject
~
\includegraphics{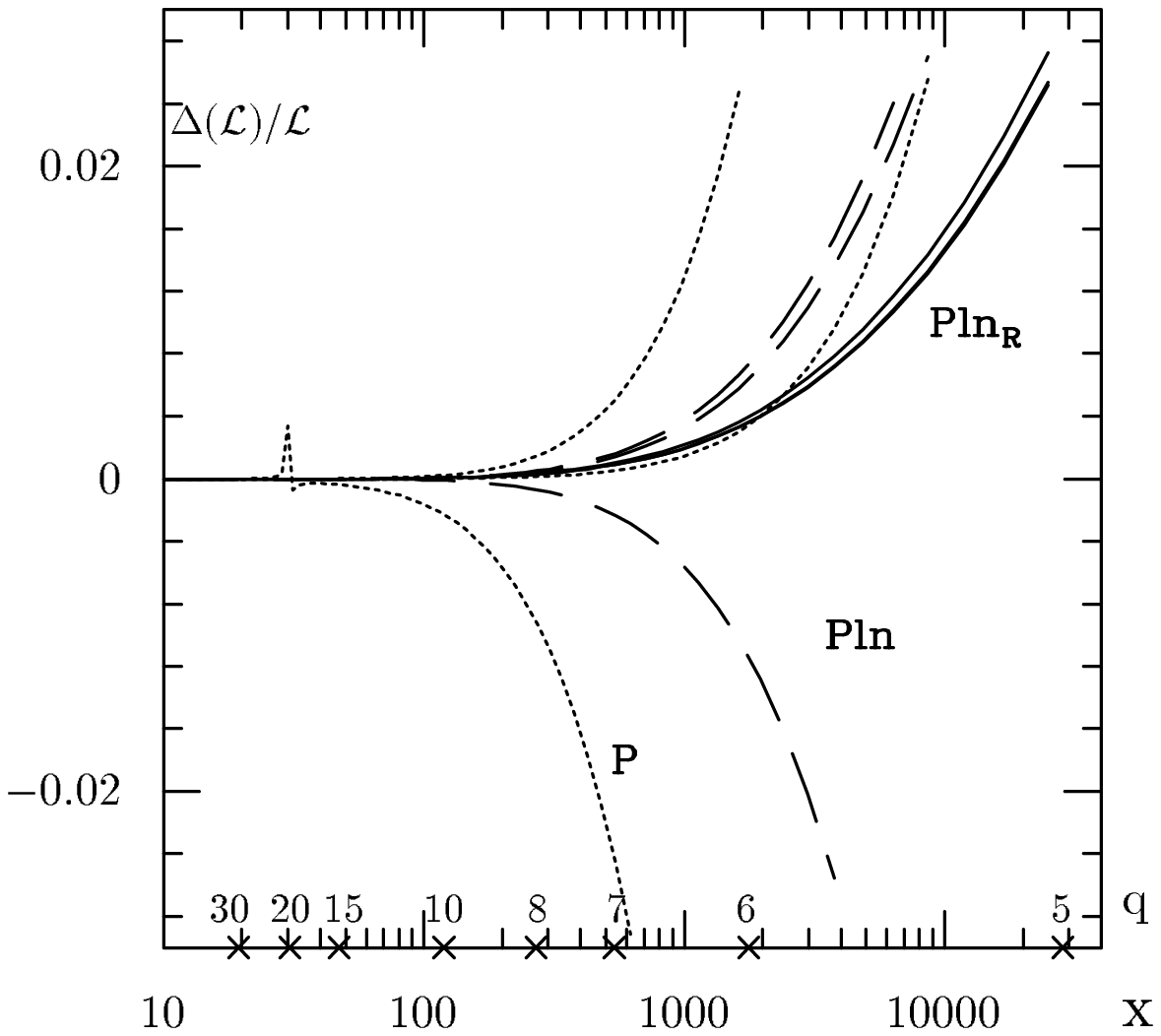}
\vskip 10.0 truecm
\centerline{\bf \fig{DL}}
\noindent {\ismallfont Relative difference to the latent heat exact result
of 4/6, 5/5 and 6/4 Pad\'e's applied to ${\cal L}^{-1}$ (dotted lines), to the
logarithmic of ${\cal L}^{-1}$ (dashed lines) and to the regularized logarithmic
(solid lines). The 4/6 simple Pad\'e has a pair of pole and zero near the real axis
giving a spurious spike near q=20.}
\bigskip

According to the above results and assumptions, we now experiment resummation
techniques by Pad\'e approximants on the series for ${\cal L}^{-1}$, which
diverges as $x^{1/2}$, \eq{Lq}. Truncating it at order 10, we want to compute
${\cal L}^{-1}$ from
$$
2 {\cal L}^{-1}=1+z+5 z^2+7 z^3+27 z^4+41 z^5+143 z^6+225 z^7+737 z^8+
  1187 z^9+3713 z^{10}+\cdots   \EQNO{LS}$$
a series which does exhibit the same qualitative aspect as $F^{(n)}, \ n\geq 2$.
In the absence of any further information on the behavior of ${\cal L}^{-1}$,
we would try a Pad\'e
resummation (P). With the knowledge we have, a better
attempt is to apply the Pad\'e resummation to the logarithm of ${\cal L}^{-1}$
(Pln). But $\log{\cal L}^{-1}\sim 1/\theta$ is still singular at $q=4$; its
 leading singularity is a pole at 1 in the variable 
$$u={2 z\over (1+\sqrt{1-4 z^2})}={2\over \sqrt{q}+\sqrt{q-4}}, \EQNO{U}$$
\vfill\eject
~
\includegraphics{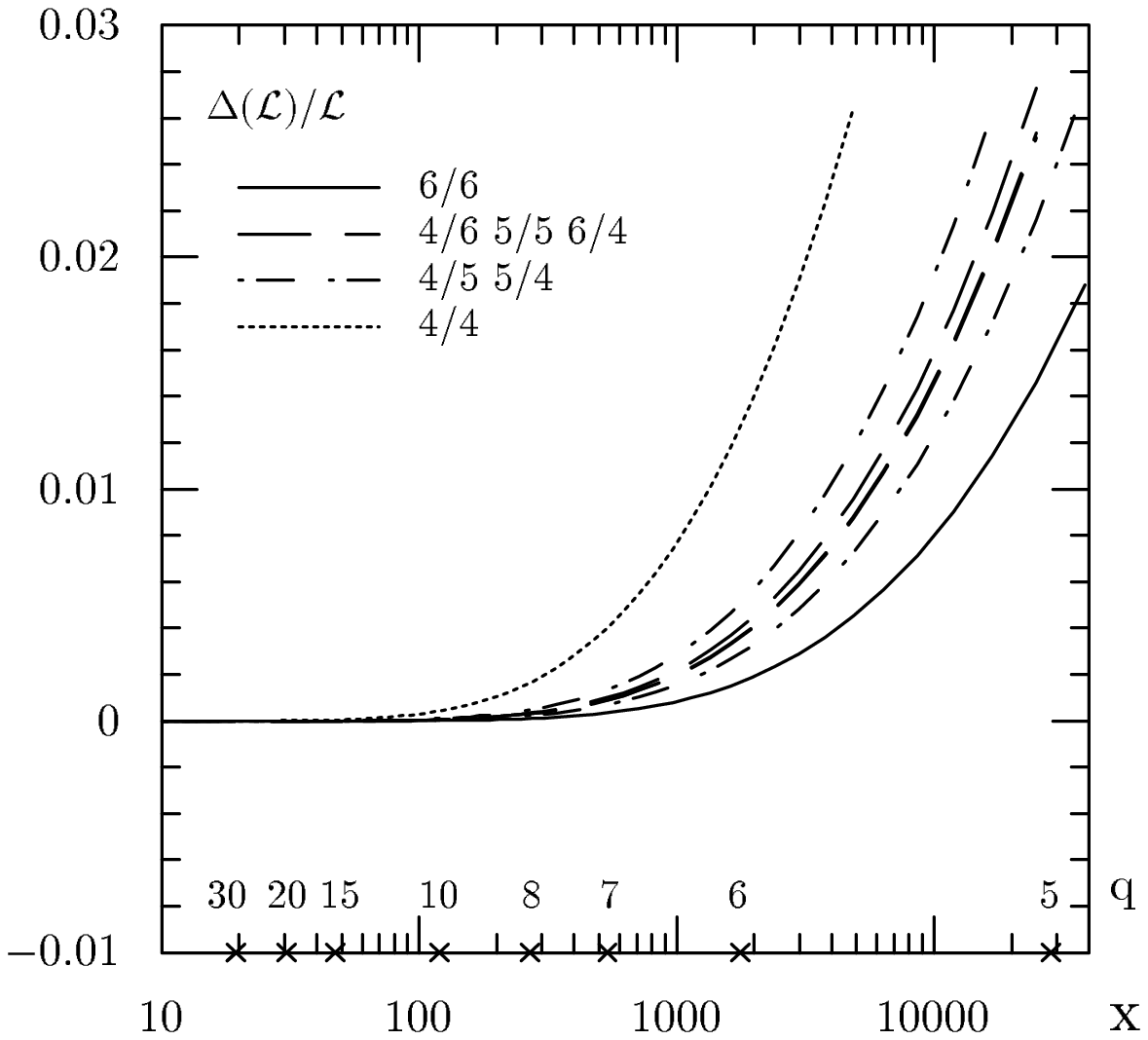}
\vskip 10.0 truecm
\centerline{\bf \fig{conver}}
\noindent{\ismallfont Relative difference to the latent heat exact result of regularized
logarithmic Pad\'e applied to truncated series at 8 (dotted), 9 (dot-dashed),
 10 (dashed) and 12 (solid line) terms.}
\bigskip
\noindent
and $u$ is expandable in $z$ around 0. So an even better Pad\'e technique in
this case is to construct Pad\'e approximants for the less singular function
$(1-u) \log(2 {\cal L}^{-1})$. Using $2 {\cal L}^{-1}$ rather than any other
constant times ${\cal L}^{-1}$ is arbitrary, but happens to be very convenient
as avoiding logarithms of numerical constants. Therefore we will construct,
for any series $S(z)$ whose lowest order term is $C_{k_{\rm min}} z^{k_{\rm min}}$ and which
is assumed to be dominantly a power of $x$ as $q\rightarrow 4$, the regularized
logarithmic Pad\'e approximant as
$$
{\rm (Pln_R)} : \ S=C_{k_{\rm min}} z^{k_{\rm min}}\exp\Bigl[{1\over 1-u}\hbox{Pad\'e} \bigl[
  (1-u)\log {S^{\rm series}\over C_{k_{\rm min}} z^{k_{\rm min}}}\bigr]\Bigr] . \EQNO{Plnr}$$

The Pad\'e $[f(z)]$'s are $P_M/Q_N$ ratios of polynomials of degrees $M$ and $N$,
their
Taylor expansion matches that of $f(z)$ up to order $M+N$. Of course, with regards
to the Pad\'e techniques here applied, a series $S$ or its inverse $S^{-1}$ leads
to the same set of approximants and the purpose of considering ${\cal L}^{-1}$
was just to illustrate its similarities with the series at hand.

The results for $M/N$ = 4/6, 5/5, and 6/4 of the techniques (P), (Pln) and
(${\rm Pln_R}$) applied to ${\cal L}$ are shown in \fig{DL}, where the error
$\Delta ({\cal L})/{\cal L}={{\cal L}^{\rm approx}-{\cal L}^{\rm exact}\over{\cal L}^{\rm exact}}$ is
plotted against the variable $x$. Typical $q$ values are also shown. As
expected, (${\rm Pln_R}$) gives the best result while (P) is worse by far
(the 4/6 approximant is even ill behaved around $q=20$). The precision reached
is still less than 3\% with (${\rm Pln_R}$) at $q$ values as ``small'' as 5.
We adopt this technique throughout the rest of this paper.  An idea of the
convergence of this resummation with the length of the series is given by
\fig{conver} showing various results for $\Delta ({\cal L})/{\cal L}$ using
the series truncated at 8, 9, 10 and 12 (the 11th term is zero).
 The convergence is fast; the diagonal
and near diagonal Pad\'e's lead to comparable results.

We now turn to the study of the $F^{(n)}$'s with $n\geq 2$ for which the series
are known to order 10 in $z$. Because their lowest term is $\sim z^2$, the
maximum value of $M+N$ available in (${\rm Pln_R}$) is 8. For $F^{(2)}$ we
find consistently that it behaves nearly linearly with $x$ for
$x\gtrsim 100 \ (q\lesssim 10)$ as well exposed by the plot of \fig{F2}

\includegraphics{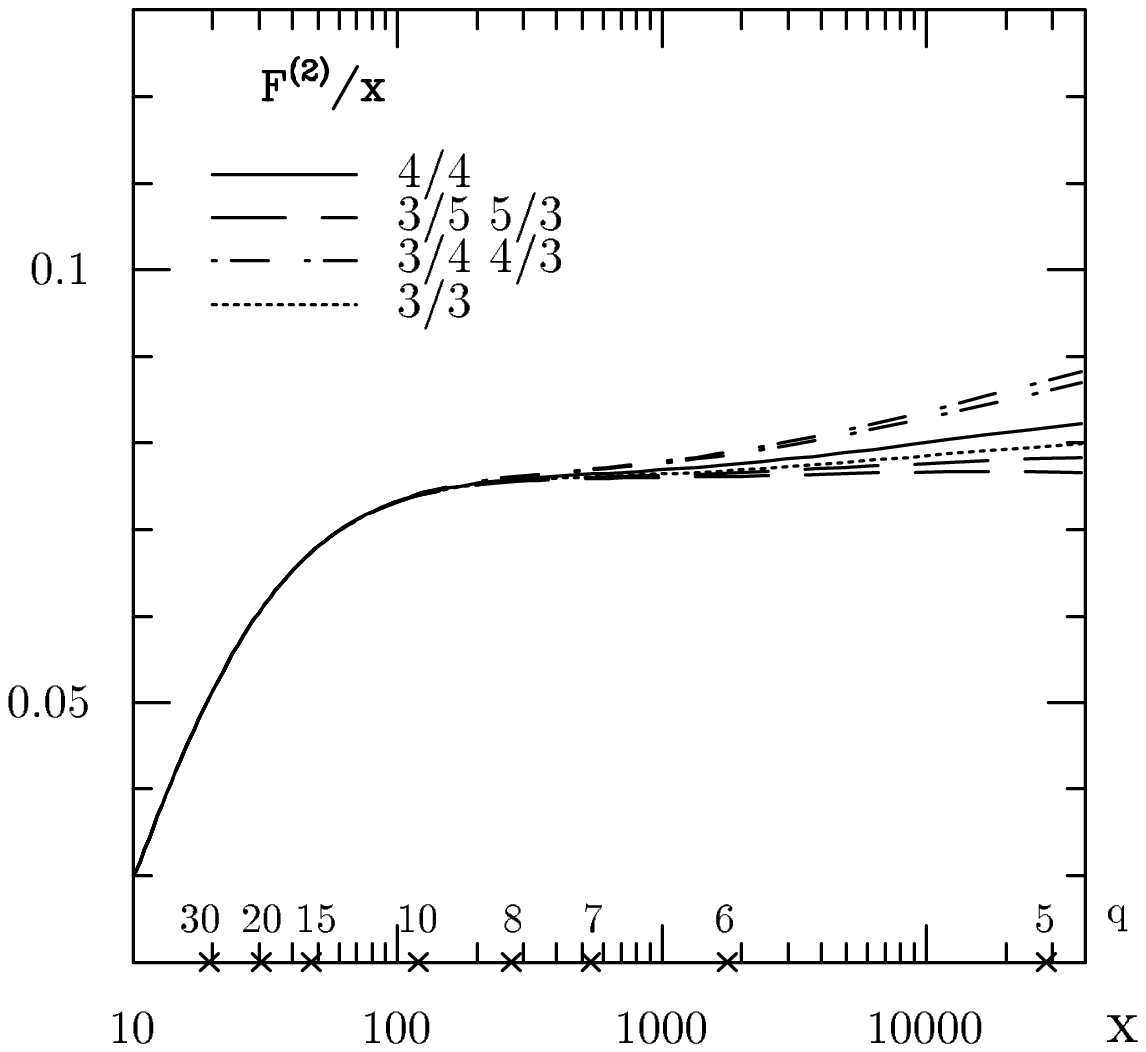} 
\vskip 10.8 truecm
\centerline{\bf \fig{F2}}
\centerline {\ismallfont Regularized logarithmic Pad\'e resummations for $F^{(2)}$.}
\vfill\eject
~
\includegraphics{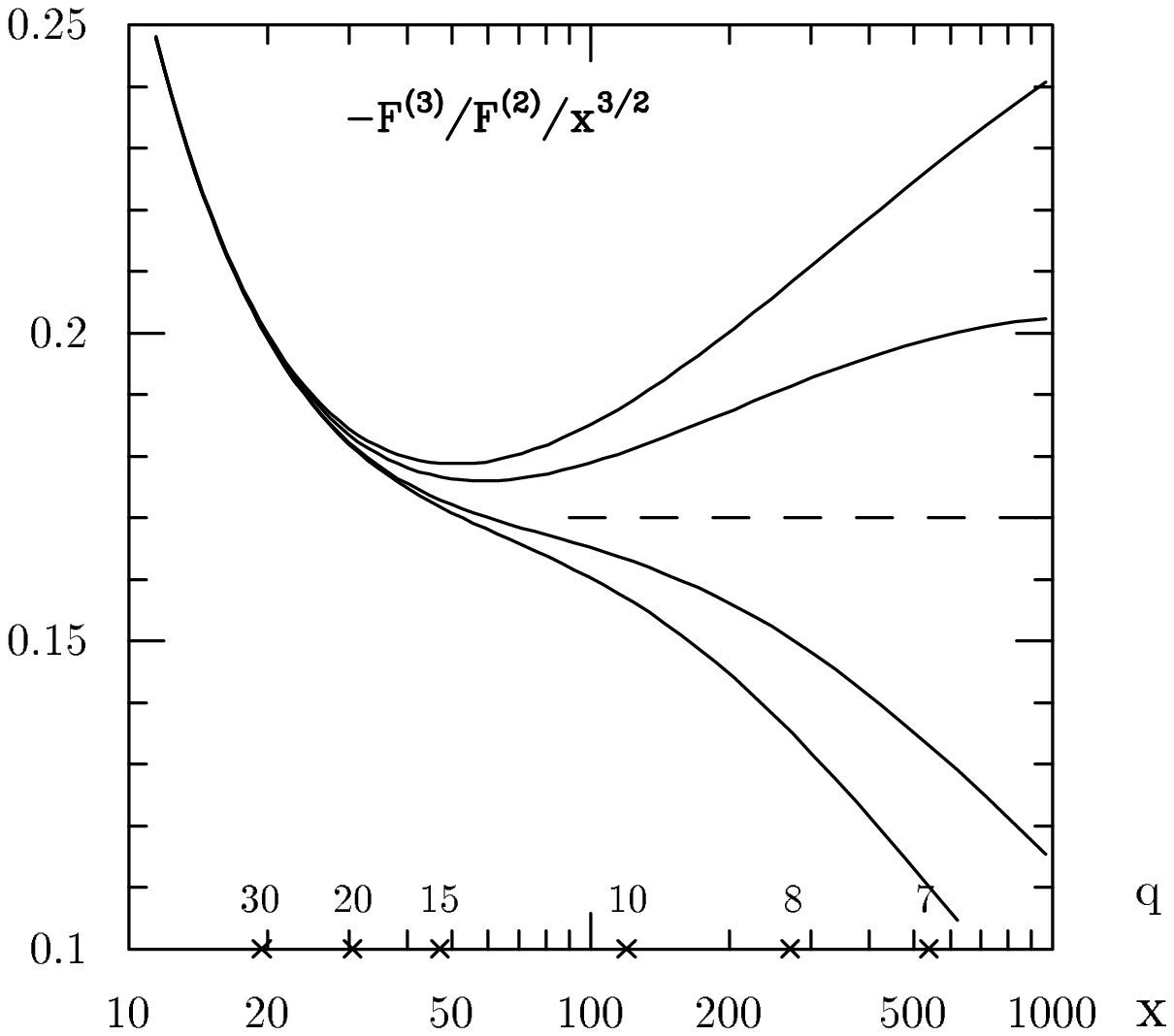} 
\vskip 10.0 truecm
\centerline{\bf \fig{F3}}
\centerline{\ismallfont Regularized logarithmic Pad\'e resummations for $F^{(3)}$.}
\bigskip 
\noindent
showing various estimates of $F^{(2)}/x$. If we {\sl assume}
that this quantity is indeed asymptotically constant, its limit is very close to
$$
F^{(2)}/x\sim \alpha =7.6 10^{-2}. \EQNO{AL}$$
In any case, there is a clear cross-over around $x=100$ between the large $q$ regime
and a very different low $q-4$ behavior. Because the various Pad\'e's lead to
very stable results for $q\geq 7$, our result also provides accurate predictions
for the specific heat (see below).

Since we have a good control on $F^{(2)}$ behavior, we next analyze $F^{(3)}$ by
the ratio $F^{(3)}/F^{(2)}$, a slightly better series than that of $F^{(3)}$ itself.
With the prejudice that $F^{(3)}\sim x^{5/2}$
describes the leading behavior and with $F^{(2)}\sim x$, it is convenient to
consider the positive quantity $-F^{(3)}/(x^{3/2}F^{(2)})$ as a function of $x$,
as shown in \fig{F3}. The four curves correspond to the Pad\'e's 3/5, 4/4, 5/3 and
6/2. Note that although $F^{(3)}$ varies by 4 orders of magnitude in the $x$
interval shown, the above ratio stays between $\sim$ .1 and .25. If now we 
{\sl assume} that $F^{(3)}/(F^{(2)}x^{3/2})$ has a finite limit at large $x$,
a reasonable estimate is
$$
{F^{(3)}\over x^{3/2}F^{(2)}} \sim \beta =.17  \EQNO{B}$$
indicated by the dashed line in \fig{F3}.

Given the estimates $\alpha$ and $\beta$ for the supposed limits of $F^{(2)}/x$
and $F^{(3)}/(x^{3/2}F^{(2)})$, the ansatz proposed in \refand{nous}{free} for
the free energy of the ordered phase gives a prediction for all the higher cumulants
$$
F^{(n)} \build{\simeq}_{x\rightarrow\infty} (-)^n\alpha \Bigl(
  {3\beta\over2}\Bigr)^{n-2}{\Gamma({2\over3}+n-2)\over\Gamma({2\over3})}
  x^{{3\over2}n-2}  \EQNO{FNAS} $$

We have compared direct ${\rm (Pln_R)}$ estimates of $F^{(n)}$, $n=2,\cdots,6$
with the above parametrization where $\alpha$ and $\beta$ have been fixed to their
guessed values. The results for $(-)^nF^{(n)}$ as a function of $x$ is shown in
\fig{Fn}. For each $n$, four curves are drawn, one showing the ansatz (dotted
lines), the three other ones resulting from 4/4, 5/3 (solid lines) and 6/2
(dashed lines) Pad\'e's. The 3/5 Pad\'e gives non-sense answers for $n$ = 5 and
6, for which case the 6/2 approximant tends to blow up at small $q$. Nothing
convincing can be extracted for $n>6$, where the series become really too short.

With this in mind, we consider the results of \fig{Fn} as
a manifest evidence for an indefinite increase with $x$ of all the
$F^{(n)}$'s,$n>1$
and a good indication that the ratios $F^{(n)}/x^{3n/2-2}$ become smooth
functions of $x$ at $x$ large enough, with relative ratios close to that of
\eq{FNAS} (note that although $F^{(6)}$ varies by $\sim 7$ orders of magnitude
when $x$ increases from 10 to 100, its value differs by less than a factor 2
from \eq{FNAS}).

Independently of any prejudice on the behavior of the cumulants, our analysis
finally provides a quantitative prediction for the values of the first cumulants
at fixed (not too small) values of $q$. Their knowledge may be of great help in
understanding better the way how the thermodynamical limit is reached in the
Potts model case, so accumulating experience on the use of finite size effects
in other studies of phase transitions. We give in Table~5, for the ordered
phase, the value of the specific heat at the transition point
($C_o=\beta^2_tF_o^{(2)}$) and of the two next moments $F^{(3)}$ and $F^{(4)}$,
together with results on $C_o$ from existing numerical simulations.
The uncertainties quoted contain some arbitrariness, as often when Pad\'e
techniques are involved. To be specific, we quote as central values the results
\vfill\eject
~
\includegraphics{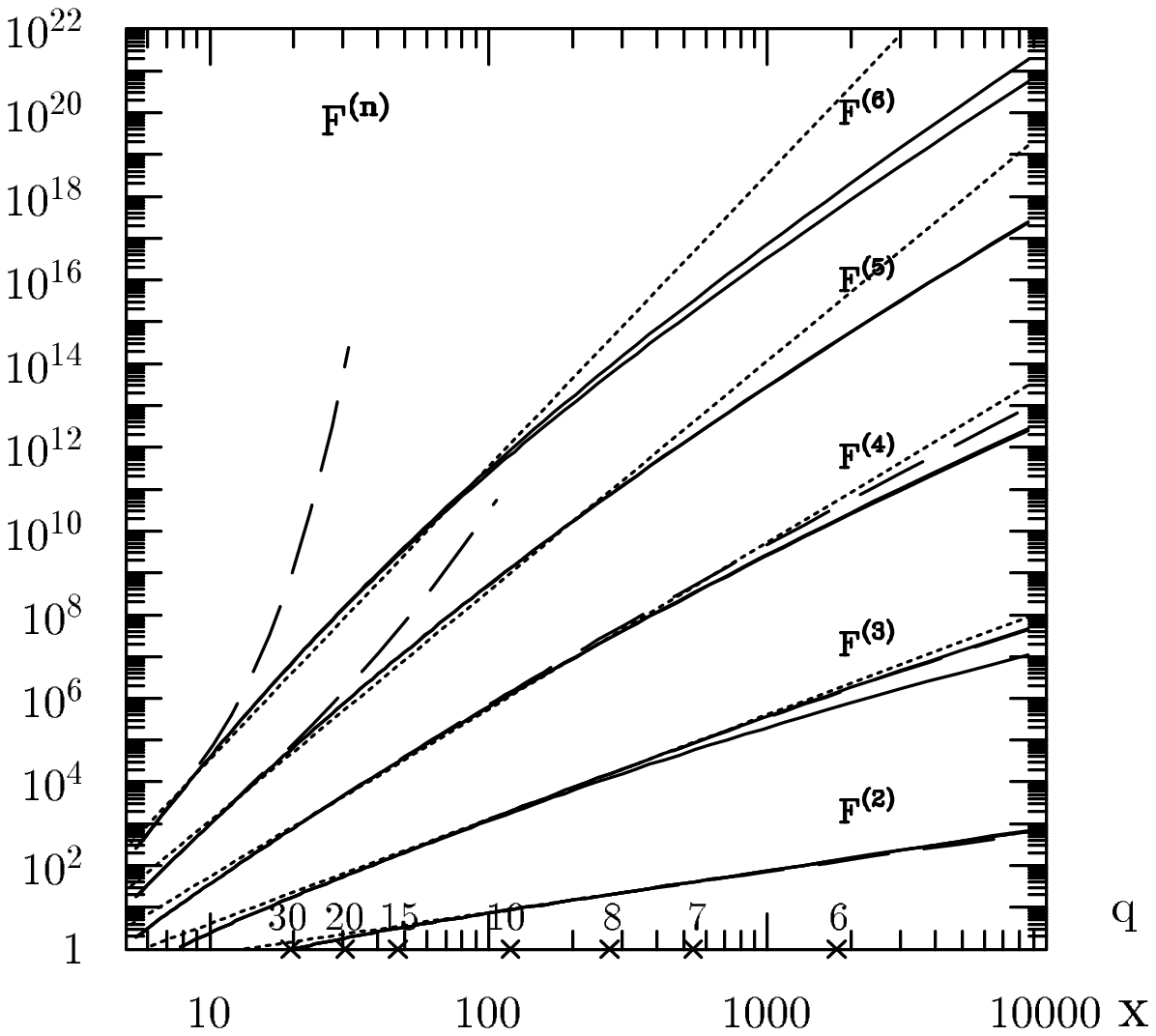} 
\vskip 10.0 truecm
\centerline{\bf \fig{Fn}}
\noindent{\ismallfont Energy cumulants $(-)^n F^{(n)}$ from 4/4, 5/3 (solid lines)
and 6/2 Pad\'e resummation and comparison to the ansatz \eq{FNAS} (dotted lines).}
\bigskip
\noindent
from 4/4 Pad\'e (for $F^{(3)}$ it is the 4/4 result of $F^{(3)}/F^{(2)}$ times
the 4/4 result of $F^{(2)}$).
The uncertainty in each case is the mean distance to the 4/3 and 3/4 results.
We clearly contradict values recently obtained from low temperature series \ref{gutt}.
\reftag{koster}\reftag{bilcom}\reftag{janke}\reftag{rummu}
\reftag{bil20}\reftag{jakap}\reftag{jakap2}
As already noticed \ref{nous}, there is strong discrepancies for $q\leq 10$
with $C_o$ values extracted from earlier numerical simulations \refto{koster}{rummu}.
Let us note that we quote two $C_o$ values from \ref{bilcom}, the value 12.7 from an
analysis at the maximum of the specific heat and the value 18. from an analysis at the
transition temperature, and this inconsistency was the starting point of the present
work.

In contrast our predictions agree very well with the most recently published values
\refto{bilcom}{jakap2}. These data correspond to higher $q$ value \ref{bilcom}
or to analysis at the transition point in the disordered phase \ref{jakap} and
in the ordered phase \ref{jakap2}. The third momentum have been measured in these
two last references and agree also very well with our predictions.

The discrepancies are observed with data obtained from FSS analysis at the
maximum of the specific heat. In fact the corresponding analysis are
neglecting  energy cumulants higher than $n=2$
in the expansions in 1/$V$ as made in \ref{koster}. 
Thus the discrepancies observed can be understood from the large values of the
higher energy cumulants as we have shown, and we can try to be more quantitative
by introducing them in the 1/$V$ expansions. For example the $\beta$
value at which the second moment is maximum expands as
$$\eqalign{
\beta_{{\rm max} F^{(2)}}& = \beta_t+{1\over V}{\ln q\over F_-^{(1)}}
 - {1\over V^2}{\ln ^2q-12\over 2}{F_-^{(2)}\over F_-^{(1)^3}}\cr
&  - {1\over V^3}{1\over 6F_-^{(1)^5}} \bigl[ 
-24 F_+^{(3)} F_-^{(1)}+(\ln^3q-36\ln q) (F_-^{(1)}F_-^{(3)}-3 F_-^{(2)^2}) \bigr]\cr
&  - {1\over V^4}{1\over 24F_-^{(1)^7}} \bigl[ 
-1152 F_-^{(1)} F_-^{(2)} F_-^{(3)}+1728 F_-^{(2)^3}\cr
&\ \ \qquad
+96 \ln q (4 F_+^{(3)}F_-^{(1)} F_-^{(2)}- F_+^{(4)}F_-^{(1)^2})\cr
&\ \ \qquad
+(\ln^4q-72\ln^2q) (F_-^{(1)^2}F_-^{(4)}-10 F_-^{(1)} F_-^{(2)} F_-^{(3)}
 +15 F_+^{(3)^3})  \bigr]\cr
+\ldots
}$$
where $F_{\pm}=F_{\rm o}\pm F_{\rm d}$ and $F_{\rm d}^{(n)}$ are obtained
from $F_{\rm o}^{(k)}$, $k=1$ to $n$,
with the duality relation. At $q=10$ we obtain
$$
\beta_{{\rm max} F^{(2)}}=\beta_t- {3.3\over V} -{2.2\over V^2}
+{1.9 \ 10^5\over V^3}-{1.5 \ 10^8\over V^4}+ \ldots
$$
It is clear that this expansion is useless in practice as well as that of
the maximum of the specific heat.

Detailed comparisons between data and consequences of our results on energy
distributions can be found in \ref{free}.

\section{Conclusions}

We have explored the properties of the 2D-Potts model free energy $F(\beta )$ in
the region $q\ge4$ where the model has a temperature driven first order transition.
This was achieved by a series expansion of $F(\beta )$, close to the transition temperature, in powers of $1/\sqrt{q}$, performed to order 10 from the Fortuin-Kasteleyn representation of the partition function.
At each order $m$ in $q^{-1/2}$, we compute the number of lattice configurations of $l$ bonds enclosing $k$ clusters of sites such that $l-2k=m$.

The results obtained were translated into similar series in $q^{-1/2}$ for the derivatives $F^{(n)}$ of the free energy taken at the transition temperature, that is for the energy cumulants of the model. These series truncated at order 10 provide strong evidence for highly divergent behaviors at a low $q$ value, the more so $n$ increases.
We assumed that this value is $q=4$, and conjectured that these singular behaviors are dominantly embodied in powers of the variable $x(q)$ of \eq{defx}, known to accurately describe the latent heat and the largest correlation length in a wide region of $q>4$.
Then, Pad\'e techniques adapted to such behaviors were applied to the $F^{(n)}$ series, leading to quantitative predictions for $n\lesssim 4\ ,\ q\gtrsim 6$, confirming severe divergences of all $F^{(n)}$'s and in semiquantitative agreement, for $n$ up to 6 at least, with an ansatz \refto{nous}{free} which prescribes the divergence rates.

First these results enlarge our analytical knowledge of the 2D-Potts model, so providing useful additional tests on methods for analyzing numerical data on finite lattices.
Next they illustrate the possible occurrence in first order transitions of properties widely influenced by a nearby continuous transition point. This may be the case as well for field driven phase transitions just below the critical temperature when the correlation is still large \refand{gupta}{rakow}.

One may think of other applications of our analytical approach to the free energy of the Potts model. An interesting ones would be a study of the analytic structure of $F_o(\beta )$ ( or $F_d(\beta )$ ) at fixed $q>4$ around $\beta =\beta_t$, where an essential singularity is expected \refand{fisher}{isakov}.
\bigskip\bigskip
We thank A.~Billoire and P.~Moussa for useful discussions and N.~Elstner for a critical reading of our manuscript.

\bigskip\centerline{\sectnfont References}\bigskip
\item{\ref{wu})} F.~Y.~Wu, Rev. Mod. Phys. {\bf 54} (1982) 235.
\item {\ref{boko})} C.~Borgs and R.~Koteck\'y, J. Stat. Phys. {\bf 61} (1990) 79;
C.~Borgs, R.~Koteck\'y and S.~Miracle-Sol\'e, J. Stat. Phys. {\bf 62} (1991) 529.
\item{\ref{xi})} A.~Kl\"umper, A.~Schadschneider and J.~Zittartz,
Z Phys. {\bf B76} (1989) 247.
\item \    E.~Buffenoir and S.~Wallon, J. Phys. {\bf A26} (1993) 3045.
\item{\ref{bojan})} C.~Borgs and W.~Janke, J. Phys. I (France) {\bf 2} (1992) 649.
\item{\ref{kast})} P.~W.~Kasteleyn and C.~M.~Fortuin, J. Phys. Soc. Japan
 {\bf 26} (Suppl.), 11 (1969).
\item{\ref{guinsp})} P. Ginsparg, Y.Y. Goldschmidt, J.-B. Zuber,
Nucl. Phys. {\bf B170} (1980) 409.
\item{\ref{enting})} I.~G.~Enting, J. Phys. A: Math. Gen. {\bf 10} (1977) 325.
\item{\ref{nous})} T.~ Bhattacharya, R.~Lacaze and A.~Morel, Europh. Lett., {\bf 23} (1993) 547.
\item{\ref{dallas})} T.~ Bhattacharya, R.~Lacaze and A.~Morel, Nucl. Phys. B (Proc. Suppl.) {\bf 34} (1994) 671.
\item{\ref{free})} T.~ Bhattacharya, R.~Lacaze and A.~Morel, Nucl. Phys. {\bf B435} (1995) 526.
\item{\ref{potts})} R.~B.~Potts, Proc. Camb. Phil. Soc. {\bf 48} (1952) 106.
\item{\ref{baxt})} R.~J.~Baxter, J. Phys. {\bf C6} (1973) L445.
\item {\ref{kopel})} J.~Hoshen and R~Kopelman, Phys. Rev. {\bf B14} (1976) 3438.
\item {\ref{gutt})}  K.~M.~Briggs, I~G~Enting and A.~J.~Guttmann, J. Phys. A: Math. Gen. {\bf 27} (1994) 1503.
\item{\ref{koster})} J.~Lee and J.~M.~Kosterlitz, Phys. Rev. {\bf B43} (1991) 3265.
\item {\ref{bilcom})} A.~Billoire, R.~Lacaze and A.~Morel,
Nucl. Phys. {\bf B 370} (1992) 773.
\item {\ref{janke})}  W.~Janke, B.~Berg and M.~Katoot, Nucl. Phys. {\bf B382} (1992) 649.
\item {\ref{rummu})}  K.~Rummukainen, Nucl. Phys. {\bf B390} (1993) 621.
\item {\ref{bil20})} A.~Billoire, T.~Neuhaus and B.~Berg, Nucl. Phys. {\bf B396} (1993) 779.
\item {\ref{jakap})}  W.~Janke and S.~Kappler, Nucl. Phys. B (Proc. Suppl.) {\bf 34} (1994) 674.
\item {\ref{jakap2})}  W.~Janke and S.~Kappler, Europhys. Lett. {\bf 31} (1995) 345.
\item {\ref{gupta})} S.~Gupta, A.~Irb\"ack, M.~Ohlsson, Nucl. Phys. {\bf B409} (1993) 663.
\item {\ref{rakow})} C.~Borgs, P.E.L.~Rakow and S.~Kappler, J. Phys. I (France) {\bf 4} (1994) 1027.
\item {\ref{fisher})} M.E.~Fisher, Physics (N.Y.) {\bf 3} (1967) 255.
\item {\ref{isakov})} S.N. Isakov, Commun. Math. Phys. {\bf 95} (1984) 427.
\vfill\eject
{\smallfont
$$\vbox{\offinterlineskip{
\def\tvi{\vrule height 8pt depth 2pt width 0pt}
\def\tv{\tvi\vrule}
\def\cc#1{\hfill\quad#1\quad\hfill}
\catcode`\*=\active \def*{\hphantom{0}}
\halign{\tv#&\cc{#}&\cc{#}&\cc{#}&\cc{#}&\cc{#}&\tv#&\cc{#}&\cc{#}&\cc{#}&\cc{#}&\cc{#}&\tv#\cr
\noalign{\hrule} 
& $k$ & $l$ &$n_{\rm plaq}$&$m$ & $N$& &$k$ & $l$ &$n_{\rm plaq}$ & $m$& $N$  &\cr
\noalign{\hrule} 
& *1& *4 & 0 & *2&****1&& 12 & 31 &*6 & *7&  ****2 &\cr
& *2& *7 & 0 & *3&****2&&    & 32 &*5 & *8&  **151 &\cr
& *3& 10 & 0 & *4&****6&&    & 33 &*4 & *9&  *2086 &\cr
& *4& 12 & 1 & *4&****1&&    & 34 &*3 & 10&  12862 &\cr
&   & 13 & 0 & *5&***18&& 13 & 34 &*6 & *8&  ***68 &\cr
& *5& 15 & 1 & *5&****8&&    & 35 &*5 & *9&  *1392 &\cr
&   & 16 & 0 & *6&***55&&    & 36 &*4 & 10&  11717 &\cr
& *6& 17 & 2 & *5&****2&& 14 & 36 &*7 & *8&  ***22 &\cr
&   & 18 & 1 & *6&***40&&    & 37 &*6 & *9&  **864 &\cr
&   & 19 & 0 & *7&**174&&    & 38 &*5 & 10&  *9332 &\cr
& *7& 20 & 2 & *6&***22&& 15 & 38 &*8 & *8&  ****6 &\cr
&   & 21 & 1 & *7&**168&&    & 39 &*7 & *9&  **456 &\cr
&   & 22 & 0 & *8&**566&&    & 40 &*6 & 10&  *7032 &\cr
& *8& 22 & 3 & *6&****6&& 16 & 40 &*9 & *8&  ****1 &\cr
&   & 23 & 2 & *7&**134&&    & 41 &*8 & *9&  **218 &\cr
&   & 24 & 1 & *8&**676&&    & 42 &*7 & 10&  *4748 &\cr
&   & 25 & 0 & *9&*1668&& 17 & 43 &*9 & *9&  ***88 &\cr
& *9& 24 & 4 & *6&****1&&    & 44 &*8 & 10&  *3010 &\cr
&   & 25 & 3 & *7&***72&& 18 & 45 &10 & *9&  ***30 &\cr
&   & 26 & 2 & *8&**656&&    & 46 &*9 & 10&  *1728 &\cr
&   & 27 & 1 & *9&*2672&& 19 & 47 &11 & *9&  ****8 &\cr
&   & 28 & 0 & 10&*6237&&    & 48 &10 & 10&  **914 &\cr
& 10& 27 & 4 & *7&***30&& 20 & 49 &12 & *9&  ****2 &\cr
&   & 28 & 3 & *8&**482&&    & 50 &11 & 10&  **426 &\cr
&   & 29 & 2 & *9&*2992&& 21 & 52 &12 & 10&  **197 &\cr
&   & 30 & 1 & 10&10376&& 22 & 54 &13 & 10&  ***68 &\cr
& 11& 29 & 5 & *7&****8&& 23 & 56 &14 & 10&  ***22 &\cr
&   & 30 & 4 & *8&**310&& 24 & 58 &15 & 10&  ****6 &\cr
&   & 31 & 3 & *9&**592&& 25 & 60 &16 & 10&  ****1 &\cr
&   & 32 & 2 & 10&13160&&   &    &   &   &        &\cr
\noalign{\hrule}
}}}$$    
}
\centerline{\bf Table 1}
\centerline{\ismallfont Number $N$ of unequivalent (upon lattice translations) dominant
configurations.}
\vfill\eject
$$\vbox{\offinterlineskip{
\def\tvi{\vrule height 10pt depth 3pt width 0pt}
\def\tv{\tvi\vrule}
\def\cc#1{\hfill \ #1 \ \hfill}
\def\cg#1{ \ #1 \ \hfill}
\catcode`\*=\active \def*{\hphantom{0}}
\halign{\tv#&\cg{#}&\tv#&\cc{#}&\tv#&\cc{#}&\tv#\cr
\noalign{\hrule}
&\hfill Contribution && Cluster set &&  &\cr
\noalign{\hrule}
&$+  2            \ C( 7,5)  $&&$      (2,7)                     $&&{\vsf 1}&\cr
&$\qquad-(2V-16)   \ C(11,1) $&&$ (2,7)\star (1,4)               $&&{\vsf 2}&\cr
&$\qquad-    12    \ C(10,2) $&&$   ((2,7)+3;3,10)               $&&{\vsf 3}&\cr
&$\qquad-     4    \ C(12,1) $&&$ ((2,7)+4;4,12)                 $&&{\vsf 4}&\cr
&$\qquad-     4               $&&$ ((2,7)+5;4,12)                $&&{\vsf 5}&\cr
&$\qquad-    28               $&&$ ((2,7)+5;4,13)                $&&{\vsf 6}&\cr
&$\qquad+     8               $&&$ (((2,7)+3;3,10)+2;4,12)       $&&{\vsf 7}&\cr
&$+    12  \ C(10,3)         $&&$(2,9;3,10)                      $&&{\vsf 8}&\cr
&$\qquad -     8  \ C(12,1)  $&&$((2,9;3,10)+2;4,12)             $&&{\vsf 9}&\cr
&$\qquad -    80              $&&$((2,9;3,10)+3;4,13)            $&&{\vsf 10}&\cr
&$+     6  \ C(12,2)         $&&$(2,10;4,12)                     $&&{\vsf 11}&\cr
&$+    48  \ C(13,1)         $&&$(2,11;4,13)                     $&&{\vsf 12}&\cr
&$+     8  \ C(15,1)         $&&$(2,11;5,15)                     $&&{\vsf 13}&\cr
&$+    80                     $&&$(2,12;5,15)                    $&&{\vsf 14}&\cr
&$+    12                     $&&$(2,12;6,17)                    $&&{\vsf 15}&\cr
&$+ 1/2(V-5)\ C( 8,4)        $&&$(1,4)\star (1,4)                $&&{\vsf 16}&\cr
&$\qquad -1/2(V^2-15V+62)     $&&$(1,4)\star (1,4)\star (1,4)    $&&{\vsf 17}&\cr
&$\qquad -  12 / 2 \ C(10,2) $&&$((1,4)\star (1,4)+2;3,10)       $&&{\vsf 18}&\cr
&$\qquad -   4/ 2             $&&$((1,4)\star (1,4)+4;4,12)      $&&{\vsf 19}&\cr
&$\qquad -  52/ 2             $&&$((1,4)\star (1,4)+4;4,13)      $&&{\vsf 20}&\cr
&$\qquad +  8 / 2          $&&$(((1,4)\star (1,4)+2;3,10)+2;4,12)$&&{\vsf 21}&\cr
&$\qquad -2/2(4V-32) \ C(11,1)$&&$(1,4)\star ((1,4)+3;2,7)       $&&{\vsf 22}&\cr
&$+(2V-16)  \ C(11,2)        $&&$(1,4)\star (1,6;2,7)            $&&{\vsf 23}&\cr
&$\qquad - 28                 $&&$((1,4)\star (1,6;2,7)+2;4,13)  $&&{\vsf 24}&\cr
&$+(6V-62)                    $&&$(1,4)\star (1,8;3,10)          $&&{\vsf 25}&\cr
&$+( V-12)                    $&&$(1,4)\star (1,8;4,12)          $&&{\vsf 26}&\cr
&$+1/2(4V-46)                 $&&$(1,6;2,7)\star (1,6;2,7)       $&&{\vsf 27}&\cr
\noalign{\hrule}
}}}$$
\centerline{\bf Table 2}
\centerline{\ismallfont Detailed contributions to $N_{2,12}/V$ in column 1 (with $C(i,j)$ as}
\centerline{\ismallfont given in \eq{defC} ) of the different cluster sets in column 2.}
\vfill\eject
{\smallfont
$$\vbox{\offinterlineskip{
\def\tvi{\vrule height 10pt depth 3pt width 0pt}
\def\tv{\tvi\vrule}
\def\cc#1{\hfill \ #1 \ \hfill}
\catcode`\*=\active \def*{\hphantom{0}}
\halign{\tv#&\cc{#}&\cc{#}&\cc{#}&\cc{#}&\cc{#}&\cc{#}&\tv#&\cc{#}&\cc{#}&\cc{#}&\cc{#}&\cc{#}&\cc{#}&\tv#\cr
\noalign{\hrule} 
&$l$&$m=$2&$m=$4&$m=$6&$m=$8&$m=$10&&$l$&$m=$1&$m=$3&$m=$5&$m=$7&$m=$9&\cr
\noalign{\hrule} 
& 60 & &      &      &        &      1    && 59 & &       &     &      &         &\cr
& 58 & &      &      &        &      6    && 57 & &      &     &      &         &\cr
& 56 & &      &      &        &     22    && 55 & &      &     &      &         &\cr
& 54 & &      &      &        &     68    && 53 & &      &     &      &         &\cr
& 52 & &      &      &        &     187   && 51 & &      &     &      &         &\cr
& 50 & &      &      &        &     328   && 49 & &      &     &      &     2   &\cr
& 48 & &      &      &        &     600   && 47 & &      &     &      &     8   &\cr
& 46 & &      &      &        &     610   && 45 & &      &     &      &    30   &\cr
& 44 & &      &      &        &       4   && 43 & &      &     &      &    88   &\cr
& 42 & &      &      &        &   -1352   && 41 & &      &     &      &   178   &\cr
& 40 & &      &      &   1    &   -2896   && 39 & &      &     &      &   252   &\cr
& 38 & &      &      &   6    &   -5198   && 37 & &      &     &      &   204   &\cr
& 36 & &      &      &  22    &    2612   && 35 & &      &     &      &  -532   &\cr
& 34 & &      &      &  68    &   -5863   && 33 & &      &     &      &  -722   &\cr
& 32 & &      &      &  89    &   24485   && 31 & &      &     &    2 & -2618   &\cr
& 30 & &      &      & 112    &  -16014   && 29 & &      &     &    8 &   620   &\cr
& 28 & &      &      &-229    &   28035   && 27 & &      &     &   30 &   894   &\cr
& 26 & &      &      &-570    &  -38351   && 25 & &      &     &   48 &  7334   &\cr
& 24 & &      &   1  &-1749/2 &  100263   && 23 & &      &     &   14 & -6054   &\cr
& 22 & &      &   6  & 1182   & -379348   && 21 & &      &     & -244 & 71360/3 &\cr
& 20 & &      &  22  &  233   & 3321646/5 && 19 & &      &     & -208 &-78920   &\cr
& 18 & &      &   6  & 6704   & -625246   && 17 & &      &   2 & -138 &106586   &\cr
& 16 & &      & -63  &-65917/4&  351774   && 15 & &      &   8 & 2156 &-226216/3&\cr
& 14 & &      &-201  & 15532  & -119493   && 13 & &      &   6 &-3000 & 29954   &\cr
& 12 & &    1 &1555/3& -7365  &   21723   && 11 & &      & -72 & 1762 & -6066   &\cr
& 10 & &    6 & -406 &  1686  &   -6431/5 &&  9 & &      &  88 & -468 &  3422/9 &\cr
&  8 & &  -33/2& 131 & -459/4 &          &&  7 & &    2 & -38 & 254/7&         &\cr
&  6 & &   12  & -37/3&        &          &&  5 & &   -4 &22/5 &      &         &\cr
&  4 & 1& -3/2 &      &        &          &&  3 & &  2/3 &     &      &         &\cr
&  2 &-1&      &      &        &          &&  1 &2&      &     &      &         &\cr
\noalign{\hrule} 
}}}$$    
}
\vskip -0.5 truecm
\centerline{\bf Table 3}
\centerline{\ismallfont Non zero coefficients $A(l,m)$ contributing to $F_o$ in \eq{Fo} }
\vfill\eject
{\smallfont
$$\vbox{\offinterlineskip{
\def\tvi{\vrule height 10pt depth 3pt width 0pt}
\def\tv{\tvi\vrule}
\def\cc#1{\hfill \ #1 \ \hfill}
\def\dd#1{\hfill #1 \ }

\catcode`\*=\active \def*{\hphantom{0}}
\halign{\tv#&\cc{#}&\tv#&\dd{#}&\dd{#}&\dd{#}&\dd{#}&\dd{#}&\dd{#}&\dd{#}&\tv#\cr
\noalign{\hrule}
&$m\setminus n$&&0&1&2&3&4&5&6&\cr
\noalign{\hrule}
&2&&    0& 4&      16&      64&        256&        1024&          4096&\cr
&3&& -4/3&  -2&    34&     430&       3778&       29518&        218914&\cr
&4&&    1&   2&   114&    2654&      41778&      556382&       6813714&\cr
&5&& -8/5&  -6&   254&   12186&     322670&     6773994&     126069374&\cr
&6&&    2&   4&   882&   57018&    2210982&    67122114&    1774583142&\cr
&7&&-12/7& -16&  1944&   224732&  12819264&   546094604&   19774354944&\cr
&8&&  5/2&   6&  6128&   888024&  68657204&  3918393456&  187361651588&\cr
&9&& -4/9& -38& 13550&  3164682& 333583598& 25037212842& 1545876302510&\cr
&10&&   0&   0& 39698& 11243178&1532324246&146961943266&11451708807878&\cr
\noalign{\hrule}
}}}$$
}
\vskip -0.5 truecm
\centerline{\bf Table 4}
\centerline{\ismallfont Coefficients $C_m^n$ of the expansion of $F_o^{(n)}$ in \eq{Fon}.}
\bigskip\bigskip\bigskip

$$\vbox{\offinterlineskip{
\def\tvi{\vrule height 10pt depth 3pt width 0pt}
\def\tv{\tvi\vrule}
\def\cc#1{\hfill \ #1 \ \hfill}
\def\vc#1{\hfill \  #1  \hfill}
\catcode`\*=\active \def*{\hphantom{0}}
\halign{\tv#&\cc{#}&\tv#&\cc{#}&\tv#&\cc{#}&\tv#&\cc{#}&\tv#&\cc{#}&\tv#&\cc{#}&\tv#&\vc{#}&\tv#\cr
\noalign{\hrule}
&$q$&&$C_o$&&$-F_o^{(3)}$&&$F_o^{(4)}$&&$C_o^{\rm exp}$&&$-F_o^{(3){\rm exp}}$&&Ref.&\cr
\noalign{\hrule}
&30&& 3.41294(5)&&16.74(2)    &&7.00(5) $10^2$&&          &&         &&             &\cr
\noalign{\hrule}
&20&& 5.3612(4) &&56.9(4)     &&5.0(1) $10^3$ && 5.2(2)   &&         && \ref{bil20} &\cr
&  &&           &&            &&              && 5.38(4)  &&55.8(9)  && \ref{jakap} &\cr
&  &&           &&            &&              && 5.351(15)&&57.0(13) && \ref{jakap2}&\cr
\noalign{\hrule}
&15&& 7.999(3)  &&179.(4)     &&3.1(2) $10^4$ && 8.04(4)  &&175(5)   && \ref{jakap} &\cr
&  &&           &&            &&              &&8.016(21) &&180.5(31)&& \ref{jakap2}&\cr
\noalign{\hrule}
&10&& 17.98(2)  &&1.9(2)$10^3$&&1.3(2) $10^6$ && 10.6(11) &&         && \ref{koster}&\cr
&  &&           &&            &&              &&12.7(3)   &&         && \ref{bilcom}&\cr
&  &&           &&            &&              &&$\sim$18. &&         && \ref{bilcom}&\cr
&  &&           &&            &&              && 18.0(2)  &&2066(81) && \ref{jakap} &\cr
&  &&           &&            &&              && 17.95(13)&&1979(87) && \ref{jakap2}&\cr
\noalign{\hrule}
&*8&& 36.9(2)   &&1.4(4)$10^4$&&2.7(8) $10^7$ &&22.8(30)  &&         && \ref{koster}&\cr
\noalign{\hrule}
&*7&& 69.6(5)   &&7.(3)$10^4$ &&3.4(13) $10^8$&& 47.5(25) &&         && \ref{janke} &\cr
&  &&           &&            &&              && 50.(10)  &&         && \ref{bilcom}&\cr
&  &&           &&            &&              && 44.(22)  &&         && \ref{rummu} &\cr
\noalign{\hrule}
}}}$$
\vskip -0.5 truecm
\centerline{\bf Table 5}
\centerline{\ismallfont Results for the first energy cumulants at some  $q$ values}
\centerline{\ismallfont and comparison to numerical data.}

\end